\documentclass[12pt,a4paper]{article}
\usepackage{amsfonts}
\usepackage[english]{babel}
\usepackage{graphics,amssymb,amsbsy,amsmath}
\usepackage[T2A]{fontenc}

\makeatletter

\renewcommand{\p@section}{}

\renewcommand{\p@subsection}{}

\newcommand{\ps@plainh}
 {\renewcommand{\@evenhead}{\hfill\textrm{\thepage}}
\renewcommand{\@oddhead}{\@evenhead}
\renewcommand{\@oddfoot}{}  \renewcommand{\@evenfoot}{} }
\makeatother
\newcommand*{\ccc} [1]{\stackrel{*}{#1}\!\!{\vphantom{#1}}}
\newcommand*{\cc}[1]{ \rlap{$\stackrel{*}{\phantom{#1}}$}#1 }
\newcommand{\uz}{{\underline z}{\vphantom{z}}}
\newcommand{\bz}{{z}}

\newcommand{\upartial}{{\underline\partial}{\vphantom{\partial}}}
\newcommand{\barM}{\hat{\bar M}{\vphantom{M}}}
\newcommand{\barMM}{\hat{\bar{\bf M}}{\vphantom{M}}^2}
\newcommand{\alphar}{{\underline{a}}}
\newcommand{\betar}{{\underline{b}}}

\newcommand{\mur}{{\underline{m}}}
\newcommand{\nur}{{\underline{n}}}
\newcommand{\ctimes}{{\times \!\!\!\!\!\!\supset}}
\newcommand{\qq}{\frac 12}
\newcommand{\diag}{\mathop{{\rm diag}}}

\newcommand{\sign}{\mathop{{\rm sign}}}

\newcommand{\da}{{\dot\alpha}}
\newcommand{\db}{{\dot\beta}}

\renewcommand{\mathbf}[1]{{\boldsymbol{#1}}}
\begin{document}

\date{}
\title{{\Large Classification of quantum relativistic orientable objects}}
\author{D.M. Gitman${}^a$\thanks{%
E-mail: gitman@dfn.if.usp.br} and A.L. Shelepin${}^b$\thanks{%
E-mail: alex@shelepin.msk.ru} \\
%EndAName
\\
${}^a$Instituto de F\'{\i}sica, Universidade de S\~{a}o Paulo,\\
Caixa Postal 66318-CEP, 05315-970 S\~{a}o Paulo, S.P., Brazil\\
${}^b${Moscow Institute of Radio Engineering, Electronics and Automation,}\\
Prospect Vernadskogo, 78, 117454, Moscow, Russia }
\maketitle

\begin{abstract}
Started from our work "Fields on the Poincar\'{e} Group and Quantum Description of
Orientable Objects" (EPJC,2009), we consider here a classification of
orientable relativistic quantum objects in $3+1$ dimensions. In such a
classification, one uses a maximal set of $10$ commuting operators
(generators of left and right transformations) in the space of functions on
the Poincar\'{e} group. In addition to usual $6$ quantum numbers related to
external symmetries (given by left generators), there appear additional
quantum numbers related to internal symmetries (given by right generators).
We believe that the proposed approach can be useful for description of
elementary spinning particles considering as orientable objects. In
particular, their classification in the framework of the approach under
consideration reproduces the usual classification but is more comprehensive.
This allows one to give a group-theoretical interpretation to some facts of
the existing phenomenological classification of known spinning particles.
\end{abstract}

\section*{Introduction}

In our previous work \cite{GitSh09}, we discussed an new approach for
description orientable quantum relativistic objects in $3+1$-dimensions. In
such an approach, the orientable object is associated with a scalar field
$f(h)$ on the Poincar\'{e} group
$M(3,1)=T(4)\ctimes {\rm Spin}(3,1)$, $h\in M(3,1)$.
The field depends on a $4$-vector $x^{\mu },$ which gives a position of the
object, and on a $6$-parameter matrix $Z\in \mathrm{Spin}(3,1)$, which
describes the object orientation.

The field $f(h)$ admits two kinds of transformations, corresponding to a
change of the laboratory, or space-fixed reference frame (s.r.f.), as well
as to a change of the local, or body-fixed reference frame (b.r.f.),
\begin{equation}
T(g_{l},g_{r})f(h)=f(g_{l}^{-1}hg_{r}).  \label{vv1}
\end{equation}%
Here left multiplication by $g_{l}^{-1}$ corresponds to a change of the
s.r.f. (Lorentz transformations), whereas right multiplication by $g_{r}$
corresponds to a change of the b.r.f.. There are two sets of transformation
generators -- right and left ones, and they are used to construct a maximal
set of commuting operators in the space of functions $f(h)$.

The set of all the transformations (\ref{vv1}) form the direct product
$M(3,1)\times M(3,1).$ Possible external symmetries correspond to the left
transformations, whereas some of possible internal symmetries correspond to
the right transformations. Indeed, external symmetries are usually defined
as symmetries of the enclosing space, i.e., symmetries with respect to a
change of the s.r.f., while internal ones are defined as symmetries of the
body itself, in particular, symmetries with respect to a change of the
b.r.f..

We believe that the proposed approach can be useful for description of
known elementary spinning particles. In particular, their
classification within this approach (considering spinning particles as
orientable objects) could be more complete and consistent.

A classification of orientable objects is natural to define with the help of
a maximal set of commuting operators, which are constructed from the
generators of transformations (\ref{vv1}). This set contains $10$ commuting
operators (according to the number of the group parameters) and consists of
$4$ operator functions of the left generators, $4$ operator functions of the
right generators and $2$ Casimir operators, which can be constructed from
the left generators, as well as from the right generators. Such a
classification attributes $10$ quantum numbers to an orientable quantum
object.

On the other side, in relativistic quantum theory of point-like objects
there exists the Wigner's classification \cite{Wig39}, based on the left
generators (generators of external symmetry transformations). Two Casimir
operators determine the representation (mass and spin), while the remaining
$4$ operators determine, for instance, helicity and momentum. Thus, the
Wigner's classification attributes only $6$ quantum numbers to a
relativistic quantum point-like object.

One ought to mention that in the 1960s, attempts were made to unite internal
and external symmetries in the framework of one group. Soon, however, the
so-called no-go theorem \cite{ColMa67} was proved (under some very general
assumptions), stating that the symmetry group of the $S$-matrix is locally
isomorphic to a direct product of the Poincar\'{e} group and the group of
internal symmetries. However, on this basis, one often makes too strong conclusion
that a nontrivial relation between internal and external symmetries is impossible.

As was already said, the transformations (\ref{vv1}) of a field $f(h)$ form
the direct product of groups of internal and external symmetries,
in agreement with mentioned no-go theorem.
Nevertheless, as will be demonstrated below, a nontrivial relation between
internal and external quantum numbers is possible. Both transformation
groups, corresponding to a change of the s.r.f. and b.r.f., act in the same
space of $10$-parameter functions $f(h)$ and have the same Casimir operators
which define the mass and the spin. By fixing eigenvalues of the Casimir
operators, and therefore fixing the representation, we obviously impose some
conditions on the spectra of both left and right operators that enter the
maximal set. Thus, in spite of the fact that the left and the right
operators commute, their spectra are not independent.

Following our work \cite{GitSh09}, studying relativistic orientable objects,
we often appeal to the intuitively clear example of a three-dimensional
rotator, described by a field on the group $SO(3)\sim SU(2)$. The left
$\hat{J}_{1},\hat{J}_{2},\hat{J}_{3}$ and right $\hat{I}_{1},\hat{I}_{2},\hat{I}_{3}$
generators of the group $SU(2)$ (being the operators of angular
momentum in the s.r.f. and b.r.f.) commute with each other and have the same
spectrum at a fixed eigenvalue of the Casimir operator
$\hat{\mathbf{J}}^{2}=\hat{J}_{1}^{2}+\hat{J}_{2}^{2}+\hat{J}_{3}^{2}=%
\hat{I}_{1}^{2}+\hat{I}_{2}^{2}+\hat{I}_{3}^{2}$.
Therefore, as long as we know the spectrum of the
operator $\hat{J}_{3}$, we also know the spectrum of the operator
$\hat{I}_{3}$, and, furthermore, these spectra must coincide.

Note that if $SU(2)$-subgroups of the direct product $SU(2)\times SU(2)$ act
in different spaces, then their Casimir operators $\hat{\mathbf{J}}^{2}$ and
$\hat{\mathbf{I}}^{2}$ (and the spectra of $\hat{J}_{3}$ and $\hat{I}_{3}$)
would be obviously independent of each other. However, in our case both
groups act in the same space (the space of functions of three coordinates of
an orientable object in the s.r.f.), which implies the equality of
$\hat{\mathbf{J}}^{2}$ and $\hat{\mathbf{I}}^{2}$.

We also note that $Z\in SL(2,C)=\mathrm{Spin}(3,1)$ is, in some sense,
redundant for a description of spin. Orientation is given by $6$ parameters,
whereas a description of spin (spin and projection) requires only $2$
parameters. There remain another $4$ quantum numbers related to the
orientation; these numbers, corresponding to right generators, are internal
ones.

The article is organized in the following way:

In section 1, we present a brief summary concerning the field on the Poincar\'{e}
group (details can be found in \cite{GitSh09}). In sections 2 and 3,
we examine two sets of commuting operators in the space of functions on the Poincar\'{e}
group that correspond to states with a fixed parity and chirality. We then consider
properties of right generators from these sets. In section 4, we consider
possible physical interpretation of given classification of orientable
objects. In sections 5 and 6, some sets of commuting operators are applied
to a classification of orientable objects with spin
$1/2$ and $1$. In section 7, we consider classification of the fields on the
homogeneous spaces of the Poincar\'{e} group.

We emphasize the fact that we examine only non-unitary finite-dimensional
representations of the group $\mathrm{Spin}(3,1)$, and, accordingly, those
of the group $M(3,1)=T(4)\ctimes\mathrm{Spin}(3,1)$,
which corresponds to finite-component relativistic wave equations
(\textquotedblleft relativistic quantum mechanics\textquotedblright\ or
\textquotedblleft one-particle sector\textquotedblright ). Consideration of
unitary representations goes beyond the scope of the present article.

\section{Orientable objects. Right and left transformations.}

As was already mentioned, for a description of orientable objects it is
convenient to use two reference frames: the laboratory (or s.r.f.,
related to the observer), with an orthobasis ${e}_{\mu }$, and the local (or
b.r.f., related to the body), with the orthobasis
${\xi}_\nur$, ${\xi}_\nur = v^\mu_{\;\;\nur}{e}_\mu$. For
Euclidean spaces, $({e}_{i},{e}_{j})=\delta _{ij}$, and thus the elements of
the matrix $V=\Vert v_{\;k}^{i}\Vert $ satisfy the condition
$\sum_{i}v_{\;k}^{i}v_{\;l}^{i}=\delta _{kl}$, that is, the matrix $V$ is
orthogonal, $V^{-1}=V^{T}$. For pseudo-Euclidean spaces (in particular, the
4-dimensional Minkowski space) the matrix $V$ is pseudo-orthogonal,
$V^{-1}=\eta V^{T}\eta $, $\eta =\diag(1,-1,\dots ,-1)$.

In Minkowski space, by using the homomorphism $SL(2,C)\sim SO_{0}(3,1)$, one
can describe the orientation by the matrix
\begin{equation}
Z=\left(
\begin{array}{cc}
z_{\;\,\underline{1}}^{1} & z_{\;\,\underline{2}}^{1} \\
z_{\;\,\underline{1}}^{2} & z_{\;\,\underline{2}}^{2}
\end{array}
\right) \in SL(2,C),  \label{par.0c}
\end{equation}
$\Xi =Z^{\dagger }EZ$, where $E=\sigma ^{\mu }{e}_{\mu }$ and
$\Xi =\sigma ^{\underline{n}}{\xi }_{\underline{n}}$.

The quantities $v^\nu_{\;\;\mur}\in SO_0(3,1)$ are expressed
in terms of $z$ \cite{GitSh09},
\begin{equation}
v^{\mu}_{\;\;\nur} = \frac{1}{2}(\sigma^\mu)_{\dot{\beta}\alpha}(\bar\sigma_\nur)^{\alphar\dot{\betar}}
z^{\alpha}_{\;\;\alphar}\cc z^{\dot\beta}_{\;\;\dot{\betar}}.
\label{reg.v0}
\end{equation}

We note that $v^{\mu}_{\;\nur}$ are tetrads, i.e., objects
transformed as vectors (with respect to the 1st index, $\mu $) under the
change of the s.r.f., being, at the same time, objects transformed as
vectors (with respect to the 2nd index, $\nur$) under change of
the b.r.f.\footnote{%
We underline \textquotedblleft right\textquotedblright\ indices in order to
avoid confusion, since we shall consider quantities at fixed values of the
indices (for instance, spinors $z_{\;\;\underline{1}}^{\alpha }$ and $z_{\;\;%
\underline{2}}^{\alpha }$).}.

The position of an orientable object in Minkowski space is therefore given
by a 4-vector $x$ (being coordinates of the origin of the b.r.f. in the
s.r.f.) and by the matrix of orientation $Z$. It is known that each 4-vector
$x$ can be associated with a hermitian $2\times 2$ matrix{\footnote{%
We use two sets of $2\times 2$ matrices $\sigma _{\mu }=(\sigma _{0},\sigma
_{k})$ and $\bar{\sigma}_{\mu }=(\sigma _{0},-\sigma _{k})$,
where $\sigma _{0}$ is a unity matrix and $\sigma _{k}$ are the Pauli
matrices,
\begin{equation}
 \sigma_0=\left(\begin{array}{cc} 1 & 0  \\ 0 & 1 \end{array}\right),\quad
 \sigma_1=\left(\begin{array}{cc} 0 & 1  \\ 1 & 0 \end{array}\right),\quad
 \sigma_2=\left(\begin{array}{cc} 0 & -i \\ i & 0 \end{array}\right),\quad
 \sigma_3=\left(\begin{array}{cc} 1 & 0 \\ 0 & -1 \end{array}\right).
 \label{par.1}
\end{equation}
}} $X$,
\begin{equation}
 X=x^\mu\sigma_\mu = \left(
 \begin{array}{cc} x^0+x^3 & x^1-ix^2 \\ x^1+ix^2 & x^0-x^3 \end{array}\right), \quad
 \det X=x_\mu x^\mu ,\quad x^\mu =\frac 12\mathop{\rm Tr}(X\bar\sigma^\mu).
 \label{par.2}
\end{equation}
Thus, the pair $(X,Z)\in M(3,1)$ uniquely determines the position and
orientation of the b.r.f. with respect to the s.r.f.; in addition a change
of the s.r.f. corresponds to left multiplication by $(A,U)^{-1}$, whereas a
change of the b.r.f. corresponds to right multiplication by
$(\underline{A},\underline{U})$:
\begin{equation}
(X',Z')=(A,U)^{-1}(X,Z)(\underline{A},\underline{U})=
(U^{-1}(X-A)(U^{\dagger})^{-1}+Z\underline{A}Z^{\dagger},\, U^{-1}Z\underline{U}),
\label{reg.1}
\end{equation}
where $A=\sigma _{\mu }a^{\mu }$ and $\underline{A}=\sigma_\mur \underline{a}^\mur$
 correspond to translations, while $U,\underline{U}\in SL(2,C)$ correspond to rotations and boosts.

Let us now consider functions of coordinates and orientation -- functions on
the Poincar\'{e} group $f(q)$, $q\in M(3,1)$. The action of the group $%
M(3,1)_{\mathrm{ext}}\times M(3,1)_{\mathrm{int}}$ (here we use the
subscripts $\mathrm{ext}$ and $\mathrm{int}$, according to the
interpretation of left transformations as external ones and that of right
transformations as internal ones) in the space of functions $f(q)$ is given
by
\begin{eqnarray}
&&\mathbb{T}(g,h)f(q)=f^{\prime}(q)=f(g^{-1}qh),  \label{reg.2} \\
&&q\leftrightarrow (X,Z),\quad g\leftrightarrow (A,U),\quad h\leftrightarrow
(\underline{A},\underline{U}).  \label{reg.3}
\end{eqnarray}
As a consequence of (\ref{reg.2}), we have
\begin{equation}
f^{\prime }(q^{\prime })=f(q),\quad q^{\prime}=gqh^{-1}.  \label{reg.3a}
\end{equation}
The mapping $q\leftrightarrow (X,Z)$ gives rise to the correspondence
\begin{eqnarray}
&&q\leftrightarrow (x,z),\quad  \hbox{where}  \quad x=(x^\mu), \; z=(z^\alpha_{\;\,\betar}),
\label{reg.7}
\\
&&\mu=0,1,2,3, \quad \alpha, b =1,2, \;
 \quad z_{\;\,\underline{1}}^1 z_{\;\,\underline{2}}^2-z_{\;\,\underline{1}}^2 z_{\;\,\underline{2}}^1=1,
\nonumber
\end{eqnarray}
and relation (\ref{reg.3a}) takes the form
\begin{equation}
f'(x',z')=f(x,z), \quad (x',z') \leftrightarrow q' = gqh^{-1}.
\label{reg.3b}
\end{equation}

Using such a parameterization, we find the following relations for left and
right transformations, corresponding to changes of s.r.f. and b.r.f.:
\begin{eqnarray}
&&T_{L}(g)f(x,z) = f(g^{-1}x,\; g^{-1}z), \;\;
  g^{-1}x \leftrightarrow U^{-1}(X-A)(U^{-1})^{\dagger}, \;\; g^{-1}z
  \leftrightarrow U^{-1}Z,               \label{gen.01}
\\
&&T_{R}(g)f(x,z) = f(xg,\;z g),\quad xg \leftrightarrow X + ZAZ^{\dagger},
  \quad z g \leftrightarrow ZU.        \label{gen.02}
\end{eqnarray}

According to (\ref{gen.01}), $x$ carries the the vector representation of
the Lorentz group, while $z$ carries the spinor representation of this
group. If one restricts the consideration to functions independent of $z$,
then (\ref{gen.01}) reduces to transformations of the left quasiregular
representation, corresponding to the case of a usual scalar field
$f'(x')=f(x)$. If one restricts the consideration to functions
independent of $x$, then (\ref{gen.01}) reduces to transformations of the
left generalized regular representation (GRR) of the Lorentz group.

Generators that correspond to translations and rotations have the form
\begin{eqnarray}
&&\hat{p}_{\mu }=i\partial /\partial x^{\mu }, \quad
 \hat{J}_{\mu\nu }= \hat{L}_{\mu\nu }+ \hat{S}_{\mu \nu},
 \label{gen.L}
\\
&&\hat{p}_{\mur }^R=-v^\nu_{\;\;\mur} \hat p_\nu, \quad
 \hat{J}_{\mur\nur }^R=\hat{S}_{\mur\nur }^R.
 \label{gen.R}
\end{eqnarray}
where $\hat{L}_{\mu \nu }=i(x_{\mu }\partial _{\nu }-x_{\nu }\partial _{\mu })$
are the operators of orbital momentum projections and $\hat{S}_{\mu \nu} $
are the operators of spin projections. The operators of right
translations can also be presented in the form ${\hat{P}}^{R}=-Z\hat{P}Z^{\dagger }$;
the operators $\hat{S}_{\mu \nu }$ and $\hat{S}^R_{\mur\nur }$
depend only on $z$ and $\partial /\partial z$,
\begin{eqnarray}
\hat S_{\mu\nu}&=&i\left(
      (\sigma_{\mu\nu})^{\;\;\beta}_\alpha z^\alpha_{\;\;\alphar} \partial_\beta^{\;\;\alphar} +
      (\bar\sigma_{\mu\nu})^\da_{\;\;\db} \cc z_\da^{\;\;\dot{\alphar}} \partial^\db_{\;\;\dot{\alphar}}\right) ,
\label{gen.SL}
\\
\hat S_{\mur\nur}^R &=&i\left(
      (\sigma_{\mur\nur})_{\;\;\betar}^\alphar z^\alpha_{\;\;\alphar} \partial_\alpha^{\;\;\betar} +
      (\bar\sigma_{\mur\nur})_{\dot\alphar}^{\;\;\dot\betar} \cc z_\da^{\;\;\dot{\alphar}} \partial^\da_{\;\;\dot{\betar}}\right) ,
\label{gen.SR}
\end{eqnarray}
where
\begin{equation}
(\sigma_{\mu\nu})_\alpha^{\;\;\beta} = \frac 14
(\sigma_\mu\bar\sigma_\nu-\sigma_\nu\bar\sigma_\mu)_\alpha^{\;\;\beta},\quad
(\bar\sigma_{\mu\nu})^\da_{\;\;\db}= \frac 14
(\bar\sigma_\mu\sigma_\nu-\bar\sigma_\nu\sigma_\mu)^\da_{\;\;\db}.
\label{sigmn}
\end{equation}

In addition, it is convenient to present the spin operators in terms of
three-dimensional vector notation, $\hat{S}_{k}=\frac{1}{2}\epsilon _{ijk}\hat{S}^{ij}$,
$\hat{B}_{k}=\hat{S}_{0k}$, see formulae (\ref{SL})--(\ref{S3B3}) of the Appendix.

Below, we also consider phase transformations of $Z$ (being symmetry
transformations for a field on the Poincar\'{e} group \cite{GitSh09}),
\begin{equation}
Z'=Ze^{i\phi},
\end{equation}
with the generator (chirality operator)
\begin{equation}
\hat\Gamma^5 = -i\partial/\partial \phi = {\textstyle \qq} \left( z^{\alpha}_{\;\;\betar} \partial_{\alpha}^{\;\;\betar} -
 \cc z_{\da}^{\;\;\dot\betar} \partial^{\da}_{\;\;\dot\betar} \right).
\label{chir}
\end{equation}
In fact, this means a transition to an analysis of the group
$M(3,1)_{\mathrm{ext}}\times M(3,1)_{\mathrm{int}}\times U(1)$.

\section{Maximal sets of commuting operators}

A maximal set of commuting operators in the space of the functions on a
group contains Casimir operators and an equal number operators that are some
functions of left and right generators \cite{BarRa77}.

Casimir operators that label irreps can be composed of both left and right
generators, so that the \textquotedblleft left\textquotedblright\ and
\textquotedblleft right\textquotedblright\ mass and spin are the same. For
the Poincar\'{e} group $M(3,1),$ we have
\begin{eqnarray}
&&\hat {\rm p}^2=\eta^{\mu\nu}\hat p_\mu \hat p_\nu = \eta^{\mur\nur}\hat p^R_\mur \hat p^R_\nur,
\label{p2}
\\
&&
 \hat {\rm W}^2 =\eta_{\mu\nu}\hat W^\mu \hat W^\nu=\eta_{\mur\nur}\hat W_R^\mur \hat W_R^\nur, \quad
\\
&&
 \hbox{where} \quad
 \hat W^\mu=\frac 12\epsilon^{\mu\nu\rho\sigma}\hat p_\nu \hat J_{\rho\sigma}
 = \qq\epsilon^{\mu\nu\rho\sigma}\hat p_\nu \hat S_{\rho\sigma},\qquad
 \hat W_R^{\underline{m}} = \frac 12\epsilon^{{\underline{m}}{\underline{n}}\underline{r}\underline{s}}
 \hat p^R_{\underline{n}} \hat S^R_{\underline{r}\underline{s}}.
\label{gen.Pcas}
\end{eqnarray}

As four operators, composed of left generators (generators of the group
$M(3,1)_{\mathrm{ext}}$), one can choose the Casimir operator
$\hat{\mathbf p}\hat{\mathbf S}=\hat{p_{k}}\hat{S^{k}}$ and the generators $\hat{p}_{k}$ of the subgroup
$M(3)_{\mathrm{ext}}$. The latter correspond to additive quantum numbers (if
the quantum number is additive, then this quantum number of composite system
is the sum of the corresponding quantum numbers of subsystems).
Eigenfunctions of these operators correspond to definite values of the
helicity and the momentum.

Functions of right generators can be chosen in different ways, for instance,
the set $\hat{\mathbf p}\mathstrut^{R}\hat{\mathbf S}\mathstrut^{R}$, $\hat p^R_\mur$ (analogous
to the set $\hat{\mathbf p}\hat{\mathbf S}$, $\hat{p}_{\mu }$ for $M(3,1)_{\mathrm{ext}}$),
corresponding to reduction
$M(3,1)_{\mathrm{int}}\supset M(3)_{\mathrm{int}}\supset T(3)_{\mathrm{int}}$.
However, as it follows from the explicit form of the operator
$\hat p^R_\mur$, their eigenfunctions $f(x,z)$
contain arbitrarily large powers of $z$ such that the corresponding
representation of the Lorentz group is infinite-dimensional.

We choose sets, corresponding to reduction
$M(3,1)_{\mathrm{int}}\supset SL(2,C)_{\mathrm{int}}$. Two Casimir operators of the
\textquotedblleft right\textquotedblright\ Lorentz group $SL(2,C)_{\mathrm{int}}$
have the form
\begin{equation}
  \hat {\mathbf S}^2 -\hat {\mathbf B}^2 =
  \frac 12 \hat S^R_{\mur\nur}\hat S_R^{\mur\nur}=
  \frac 12 \hat S_{\mu\nu}\hat S^{\mu\nu}, \quad
 \hat{\mathbf S}\hat{\mathbf B} =
 \frac 1{16}\epsilon^{\mur\nur\underline r\underline s}
 \hat S^R_{\mur\nur}\hat S^R_{\underline r\underline s} =
 \frac 1{16}\epsilon^{\mu\nu\rho\sigma}\hat S_{\mu\nu}\hat S_{\rho\sigma}.
\label{gen.Lcas}
\end{equation}
where $\hat{S}^{i}$ and $\hat{B}^{i}$ are operators of spin and boost
projections, see Appendix. In contrast to $\hat{S}_{\mur\nur }^R$, being
the generators of the group $M(3,1)_{\mathrm{int}}$, the operators of spin
projections $\hat{S}_{\mu \nu }$ are not generators of the group
$M(3,1)_{\mathrm{ext}}$, see (\ref{gen.L}) and (\ref{gen.R}), and therefore
operators (\ref{gen.Lcas}) are functions of right (but not left) generators
of the Poincar\'{e} group.

For $SL(2,C)_{\mathrm{int}}$ we use two sets of commuting operators
corresponding to the reduction schemes
\begin{eqnarray*}
&&SL(2,C)_{\mathrm{int}}\supset U(1)\times U(1), \\
&&SL(2,C)_{\mathrm{int}}\supset SU(2)\supset U(1).
\end{eqnarray*}
In the first case two generators $\hat{S}_{3}^{R}$ and $\hat{B}_{3}^{R}$ of
the maximal commutative (Cartan) subgroup of $SL(2,C)_{\mathrm{int}}$
correspond to additive quantum numbers. In the second case the set includes
Casimir operator $\hat{\mathbf{S}}_{R}^{2}$ of $SU(2)_{\mathrm{int}}$ and $%
\hat{S}_{3}^{R}$.

Therefore, we shall consider two sets of 10 commuting operators on the group
$M(3,1)$:
\begin{eqnarray}
&&\hat{\mathrm{W}}^{2},\;\hat{p}_{\mu },\;\hat{\mathbf{p}}\hat{\mathbf{S}}\;
(\hat{S}_{3}\,\,\hbox{in the rest frame}),\;\hat{\mathbf{S}}^{2}-\hat{\mathbf{B}}^{2},\;
\hat{\mathbf{S}}\hat{\mathbf{B}},\;\;\hat{S}_{3}^{R},\;\hat{B}_{3}^{R},
\label{31set}
\\
&&\hat{\mathrm{W}}^{2},\;\hat{p}_{\mu },\;\hat{\mathbf{p}}\hat{\mathbf{S}}\;
(\hat{S}_{3}\,\,\hbox{in the rest frame}),\;\hat{\mathbf{S}}^{2}-\hat{\mathbf{B}}^{2},\;
\hat{\mathbf{S}}\hat{\mathbf{B}},\;\;\hat{\mathbf{S}}_{R}^{2},\;\hat{S}_{3}^{R},
\label{31set-S}
\end{eqnarray}
including the Lubanski--Pauli operator $\hat{W}^{2}$, four left generators
$\hat{p}_{\mu }$ (the eigenvalue of the Casimir operator $\hat{\mathrm{p}}^{2}$,
is evidently expressed through their eigenvalues), and helicity
$\hat{\mathbf{p}}\hat{\mathbf{S}}$, expressed through the left generators.
The Casimir operators $\hat{\mathbf{S}}^{2}-\hat{\mathbf{B}}^{2}$ and
$\hat{\mathbf{S}}\hat{\mathbf{B}}$ of the subgroup $SL(2,C)_{\mathrm{int}}$
determine characteristics $j_{1},j_{2}$ of the irreps $T_{[{j_{1}j_{2}}]}$
of the Lorentz group (see Appendix).

Eigenfunctions of the maximal sets of operators (\ref{31set}) are at the
same time eigenfunctions of the chirality operator $\hat{\Gamma}^{5}$
(\ref{chir}). Indeed, in the irreps of the Lorentz group $T_{[j_{1}j_{2}]}$
an eigenvalue of the chirality operator is $\Gamma ^{5}=j_{1}-j_{2}$.

Besides the states of definite chirality, the states of definite internal
parity are obviously also of interest. The operator of space reflection $\hat P$
anticommutes with the chirality operator $\hat{\Gamma}^{5}$, and also with
the operators $\hat{\mathbf{S}}\hat{\mathbf{B}}$ and $\hat{B}_{3}^{R}$.
Therefore, eigenfunctions of these the latter three operators change their
sign under the action of $\hat P$. The set (\ref{31set-S}) is more convenient to
describe states of definite internal parity, because only one operator
$\hat{\mathbf{S}}\hat{\mathbf{B}}$ from this set doesn't commute with $\hat P$.
Eigenvalues of $\hat{\mathbf{S}}\hat{\mathbf{B}}$ are proportional to
$(j_{1}-j_{2})(j_{1}+j_{2}+1)$, see (\ref{Lcas1}). Thus, one can use quantum
numbers, corresponding to the set (\ref{31set-S}), to characterize
eigenstates of $\hat P$, with only one change (replacement the sign of
$j_{1}-j_{2}$ by internal parity $\eta $).

\section{Right generators and charges. Spectra}

Despite the commutativity of the left and right generators, the
corresponding quantum numbers are not independent. Since the same Casimir
operators are constructed from left and right generators uniformly, within
the framework of a representation (fixing their eigenvalues), the spectra of
the corresponding left and right generators are the same.

The relation between left and right generators can be easily seen on the
example of the group of three-dimensional rotations $SU(2)$, describing a
three-dimensional non-relativistic rotator \cite{Wigne59,LanLi3,BieLo81}.
Until recently, it has been the only example of a well-developed (by Wigner,
Casimir and Eckart, back in the 1930's) theory, based on use of both left
and right transformations.

The concept of two coordinate systems is always present in the problem of
rotation of a solid body, independently of the fact whether it is described
classically or quantum mechanically. One coordinate system (laboratory, or
s.r.f.) is associated with the surrounding objects, while another one
(molecular, or b.r.f.) is associated with the body. Accordingly, there are
two sets of operators of angular momentum -- in the s.r.f. (left generators
of the rotation group $\hat{J}_{k}$) and in the b.r.f. (right generators of
the rotation group $\hat{I}_{k}$).

The maximal set of commuting operators in the space of functions on the
group $SU(2)\sim SO(3)$ consist of total angular momentum
$\hat{\mathbf{J}}^{2}=\hat{\mathbf{I}}^{2}$ and two projections: $\hat{J}_{3}$
in s.r.f. and $\hat{I}_{3}$ in b.r.f. A classification of rotator sates
$|j\,m\,k\>$ is made with the help of this set,
\begin{equation}
\hat{\mathbf{J}}^{2}|J\,m\,k\>=j(j+1)|J\,m\,k\>,\quad
\hat{J}_{3}|J\,m\,k\>=m|J\,m\,k\>,\quad
\hat{I}_{3}|J\,m\,k\>=k|J\,m\,k\>.
\label{rot.1a}
\end{equation}
By virtue of the relation $\hat{\mathbf{J}}^{2}=\hat{\mathbf{I}}^{2}$ the
left and right irreps are labeled by the same $j$ and the operators $\hat{J}_{3}$
and $\hat{I}_{3}$ have the same spectrum, namely, their eigenvalues $m$
and $k$ belong to the set $-j,-j+1,\ldots ,j-1,j$. The operator $\hat{I}_{3}$,
distinguishing equivalent representations in the decomposition of the left
GRR of the rotation group into irreps, corresponds to an additive quantum
number, independent of a choice of the s.r.f. This number plays an important
role in the theory of molecular spectra \cite{BieLo81,Zare88}.

Let us now consider the groups of motions $M(D)$ and $M(D,1)$, including
rotations and translations. The generators of left rotations in this case
consist of two summands -- the orbital and spin momenta,
$\hat J_{\mu\nu}=\hat L_{\mu\nu}+\hat S_{\mu\nu}$, whereas the generators
of right rotations depend only on $z$,
$\hat J^R_{\mur\nur}=\hat S^R_{\mur\nur}$.

The simplest example is a three-parameter group $M(2)$. Here, we deal only
with one operator of projection of the angular momentum, $\hat J=\hat L+\hat S$;
the operator of right projection coincides with the operator of spin
projection, $\hat J^R=-\hat S$. The maximal set of commuting operators can
be composed of the operators of momentum and spin:
\begin{equation*}
\hat p_k, \hat S.
\end{equation*}

For the group $M(3)$, the maximal set contains 6 commuting operators, which
can be chosen as (reduction $M(3)_{\mathrm{ext}}\supset T(3)_{\mathrm{ext}}$,
$M(3)_{\mathrm{int}}\supset \mathrm{Spin}(3)_{\mathrm{int}}\supset U(1)_{\mathrm{int}}$)
\begin{equation}
\hat p_k,\; \hat p_k \hat J^k = \hat p_k \hat S^k,\;
\hat S_k \hat S^k = \hat S^R_k \hat S^{Rk},\; \hat S^R_3.
\end{equation}

Functions on the group $M(3,1)$ depend on 10 parameters, and, accordingly,
there are 10 commuting operators (two Casimir operators and two sets of four
operators, constructed from left (\ref{gen.L}) and right (\ref{gen.R})
generators), see (\ref{31set}).

For a fixed mass and spin, i.e., in the framework of a representation
determined by the Casimir operators of the Poincar\'{e} group, the spectra
of left and right generators prove to be the same. In a similar way, the
Casimir operators of the Lorentz group (\ref{gen.Lcas}) are constructed from
$\hat{S}^{\mu \nu }$ or $\hat S^{\mur\nur}_R$
uniformly, and, therefore, the spectrum of operators of left and right spin
projections for fixed values of $j_{1}$ and $j_{2}$ is the same. In
particular, Casimir operators $\hat{\mathbf{S}}^{2}=\sum (S_{k})^{2}$ and
$\hat{\mathbf{S}}_{R}^{2}=\sum (S_{k}^{R})^{2}$ of compact subgroups
$SU(2)_{\mathrm{ext}}\supset \mathrm{Spin}(3,1)_{\mathrm{ext}}$ and
$SU(2)_{\mathrm{int}}\supset \mathrm{Spin}(3,1)_{\mathrm{int}}$ with eigenvalues
$S(S+1)$ and $S_{R}(S_{R}+1)$ have the same spectrum; $S$ and $S_{R}$ belong to
the set $|j_{1}-j_{2}|,\,|j_{1}-j_{2}|+1,\dots ,j_{1}+j_{2}$, see (\ref{reduc1}).
Note that \textquotedblleft right\textquotedblright\ quantum numbers
$S^{R},\;S_{3}^{R},\;B_{3}^{R}$ can be only integer for particles of integer
spin and half-integer for particles of half-integer spin.

Left generators of the Poincar\'{e} group $\hat p_\mu$ and $\hat J_{\mu\nu}$
are associated with additive quantum numbers -- the momentum and total
angular momentum projections. Right generators $\hat p_\mu^R$ and
$\hat J_{\mu\nu}^R$, and, in particular, the operators $\hat B^R_3$ and
$\hat S^R_3 $ entering the maximal set also are associated with additive
quantum numbers.

Right generators commute with left ones and, consequently, with the
corresponding finite transformations. Therefore, under finite left
transformations (changes of the s.r.f.) the eigenfunctions of right
generators remain eigenfunctions with the same eigenvalues. In other words,
right generators determine internal quantum numbers that do not
change under changes of the s.r.f. They can be identified with charges.
Indeed, charges are usually understood as additive numbers that do not
change under changes of the s.r.f. (Lorentz transformations). Right
generators from the commutative (Cartan) subgroup satisfy such a definition.

Therefore, an orientable object is characterized by 10 quantum numbers -- 6
numbers (momentum $p_{\mu }$, spin, helicity) are determined by the left
generators and Casimir operators, whereas $4$ numbers are determined by the
right generators (Lorentz characteristics $j_{1},j_{2}$ and two charges). In
comparison with the usual description of fields in Minkowski space
($4$-momentum, spin, spin projection, representation of the Lorentz group),
there appear two additional quantum numbers.

However, there is an essential difference between left and right generators:
whereas left generators correspond to external, exact, symmetries, right
generators correspond to internal symmetries which may be broken.

Let us turn to the example of a nonrelativistic rotator, described by the
quadratic Hamiltonian
\begin{equation}  \label{rot.11}
\hat H = \sum A_k(\hat I_k)^2,
\end{equation}
where $A_k$ are the moments of inertia. For a completely symmetric rotator
($A_1=A_2=A_3=A$), not only left transformations but also right ones are
symmetry transformations of Hamiltonian (\ref{rot.11}); the symmetry group
is $SO(3)_{\mathrm{ext}}\times SO(3)_{\mathrm{int}}$. In the axially
symmetric case, only the right rotations (with the generator $\hat I_3$)
about the axis $\boldsymbol{\xi}_3$ is a symmetry of the body; the symmetry
group is $SO(3)_{\mathrm{ext}}\times SO(2)_{\mathrm{int}}$. This symmetry
corresponds to the additive quantum number $k$ (see (\ref{rot.1a})).
Finally, in the case of three different momenta of inertia the body is
asymmetric, and therefore the left transformations with the generators
$\hat I_k$ are not its symmetries; the symmetry group is $SO(3)_{\mathrm{ext}}$.

Returning to the 3+1-dimensional case, we consider the operator
$\hat p_\mu \hat \Gamma^{\mu \underline 0}$, where $\Gamma^{\mu \underline 0}$
are linear differential operators in $z$; this operator is invariant under the
transformations
$M(3,1)_{\mathrm{ext}}\times\mathop{{\rm Spin}}(3)_{\mathrm{int}}\times U(1)$
\cite{GitSh09}. The equations for the eigenvalues of this operator
\begin{equation}  \label{fseq}
\hat p_\mu \hat \Gamma^{\mu \underline 0}f(x,z)= \varepsilon ms f(x,z),
\end{equation}
where $\varepsilon =\pm 1$, in the subspaces
$f(x,z,\cc\uz)$ $f(x,\uz,\cc z)$, after a separation of the spatial and
orientation variables for spin 1/2 and 1 (corresponding to polynomials of
1st and 2nd degree with respect to $z$), turn into the equations of Dirac
and Duffin--Kemmer (see \cite{GitSh09} for details).

The chirality operator $\hat \Gamma^5$, and also the operators
$\hat{{\mathbf{S}}}\hat{{\mathbf{B}}}$ and $\hat B_3^R$ from the set (\ref{31set})
don't commute with both space reflection $\hat P$ and
$\hat p_\mu \hat \Gamma^{\mu \underline 0}$. Thus the operator $\hat B_3^R$, as well as
$\hat \Gamma^5$, corresponds to a broken symmetry. In the subspace of functions
$f(x,z,\cc\uz)$ we have $\hat\Gamma^5=i\hat B^R_3$, whereas in the subspace
$f(x,\cc z,\uz)$ we have $\hat\Gamma^5=-i\hat B^R_3$.
(Here, the multiplier $i$ originates from the fact that we consider
non-unitary finite-dimensional representation of the Lorentz group, for
which the boost generators are anti-hermitian.) Like chirality, $B^R_3$ is
an additive conserved number only for massless particles. Massive particles
correspond to eigenfunctions of the operator of space parity $\hat P=\hat I_s$ and the
operator $\hat p_\mu \hat \Gamma^{\mu \underline 0}$, which do not commute
with the operators $\hat\Gamma^5$ and $\hat B^R_3$. Therefore, massive
particles, described by equations of the form (\ref{fseq}), cannot have a
definite charge $B^R_3$ and chirality $\Gamma^5$.

The quantum number corresponding to the generator $\hat S^R_3$ must be
conserved also for massive Dirac particles, since $\hat S^R_3$ commutes not
only with all left generators of the Poincar\'{e} group, but also with the
operators $\hat p_\mu \hat \Gamma^{\mu \underline 0}$ and $\hat P$.

The question arises whether the known elementary particles possess a
conserved quantum number corresponding to the operator $\hat S^R_3$. An
answer to this question, in fact, will also answer another question, whether
one can consider usual particles as ``orientable objects'' in the sense of
the above definition.

\section{$S_3^R$ charge}

One can attempt to associate the charge corresponding to the right generator
$\hat{S}_{3}^{R}$ of the Poincar\'{e} group to observable characteristics of
physical particles. Notice that although the \textquotedblleft
internal\textquotedblright\ quantum numbers corresponding to right
generators do not change under left transformations, the discrete
transformations (automorphisms of the Poincar\'{e} group) act on both right
and left generators (see \cite{GitSh09} for details). The known behavior
under discrete transformations helps one to identify right characteristics
with properties of physical particles.

As to the operator $\hat{S}_{3}^{R}$:

1. It corresponds to an additive (conserved) quantum number.

2. It has integer eigenvalues for particles of integer spin and half-integer
eigenvalues for particles of half-integer spin.

3. It does not change sign under space reflection.

4. It changes sign under charge conjugation.

Then it follows that this number equals to zero for real neutral particle
(i.e., particle that coincides its own antiparticle); at the same time real
neutral particles with definite $S_{3}^{R}$ have to have integer spin.

If one considers $S_{3}^{R}$-charge of particles described by
finite-dimensional representations of the Lorentz group $T_{[j_{1},j_{2}]}$,
$j_{1}+j_{2}=s$, then for particles of spin $1/2$ two values are possible:
$1/2$ and $-1/2$; for particles of spin 1 three values are possible: $1,0$
and $-1$. In particular, for a photon and $Z^{0}$-boson, as real neutral
particles, we have $S_{3}^{R}=0$.

Let us see which values of $S_{3}^{R}$ can be associated with particles with
wave functions $f(x,z)$ that are eigenfunctions of the operator $\hat{S}_{3}^{R}$,
$\hat{S}_{3}^{R}f(x,z)=S_{3}^{R}f(x,z)$. Since the sign of $S_{3}^{R}$
changes under the charge conjugation, particles and antiparticles
must have $S_{3}^{R}$-charges of opposite signs. For definiteness, let an
electron $e^{-}$ has the $S_{3}^{R}$-charge $-\frac{1}{2}$; then a positron
$e^{+}$ has $S_{3}^{R}=\frac{1}{2}$. For $W^{-}$, as a charged particle, it
is natural to expect $S_{3}^{R}=\pm 1$ (the value $S_{3}^{R}=0$ is excluded
by a more detailed analysis, see below). Next, since $\tilde{\nu}_{e}$ only
admits the values $\pm \frac{1}{2}$, the reaction
$W^{-}\rightarrow e^{-}+\tilde{\nu}_{e}$ implies $S_{3}^{R}=-\frac{1}{2}$
for $\tilde{\nu}_{e}$ and $S_{3}^{R}=-1$ for $W^{-}$. Therefore, we have
\begin{equation}
\begin{array}{rcccrcccc}
 \frac{1}{2}: & \nu _{e} & e^{+} & \quad & 1: & W^{+} & \quad & 0: & \gamma ,Z^{0} \\
-\frac{1}{2}: & e^{-} & \tilde{\nu}_{e} &  & -1: & W^{-} &  &  &
\end{array}
\label{s3rzar1}
\end{equation}

Applying the same consideration to other families of fundamental fermions,
we find the following classification with respect to the sign of $S_{3}^{R}$
\begin{equation}
\begin{array}{rccccccc}
 \frac{1}{2}: & \nu _{e} & \nu _{\mu } & \nu _{\tau } & \; & u & c & t \\
-\frac{1}{2}: & e^{-} & \mu & \tau &  & d & s & b%
\end{array}
\label{s3rzar2}
\end{equation}
Therefore, the $S_{3}^{R}$-charge, whose sign changes under both
$\hat C\hat P\hat T$-transformation and charge conjugation $\hat C$, distinguishes
not only particles and antiparticles but also the \textquotedblleft up-down\textquotedblright\
components in doublets of elementary fermions. This charge is conserved in
any interactions, since the carriers of electromagnetic and strong
interactions are characterized by $S_{3}^{R}=0$, whereas we have already
examined weak charged currents.

As a consequence of (\ref{s3rzar1}) and (\ref{s3rzar2}), we find the
following empirical expression for $S_{3}^{R}$ in terms of other charges:
\begin{equation}
S_{3}^{R}=\frac{L-B}{2}+Q,  \label{s3rLBQ}
\end{equation}
where $L,B,Q$ are the lepton, baryon and electric charges, respectively.
This formula relates the \textquotedblleft right\textquotedblright\ charge
$S_{3}^{R}$ with observable characteristics of particles.

For the above-mentioned fundamental particles of spin 1 the charge
$S_{3}^{R} $ coincides with the electric charge, whereas for spin $1/2$
particles it coincides with the mean value of the electric charge
corresponding to a lepton or quark doublet.

We also note that for left particles and right antiparticles $S_{3}^{R}$
coincides with the projection of the weak isospin $T_{3}$.

Let us now consider the relation between \textquotedblleft
right\textquotedblright\ quantum number $S_{3}^{R}$ and spin. We have
already noted that spectra of right and left spin operators are not
independent, in particular, $S_3^R$ can be only integer for particles of
integer spin and half-integer for particles of half-integer spin $s$,
\begin{equation*}
(-1)^{2S_{3}^{R}}=(-1)^{2s},
\end{equation*}%
and due to (\ref{s3rLBQ}) we have
\begin{equation}
(-1)^{L-B+2Q}=(-1)^{2s}.
\end{equation}

In 1961 Michel and Lur\c{c}at \cite{LurMi61} have noted that for all the
known particles with integer $B$ there holds the relation
\begin{equation}
(-1)^{B+L}=(-1)^{2s},  \label{s3.BL}
\end{equation}
in other words, $B+L+2s$ is always even. Later, this observation resulted in
the concept of $R$-parity, being positive for all the known particles, and
defined as
\begin{equation}
R=(-1)^{3(B-L)+2s}
\end{equation}
or $R=(-1)^{3B+L+2s}$, where the multiplier $3$ is introduced for an
inclusion of quarks into the analysis.

On the condition that the electric charge is integer, relation (\ref{s3.BL})
is a consequence of (\ref{s3rLBQ}). Indeed, since the right projection
$S_{3}^{R}$ and spin $s$ must take integer and half-integer values
simultaneously, both $(B\pm L)/2$ and spin $s$ are integer and half-integer
simultaneously. Furthermore, for fractional $1/3$-multiple charges (quarks),
we have $(-1)^{2s}=(-1)^{2S_{3}^{R}}=(-1)^{6S_{3}^{R}}=(-1)^{3(L-B)}$, since
$6Q$ is even-valued. As a consequence, for all particles with integer charge
$Q$ or with $1/3$-multiple charge $Q$, $R$-parity is positive.

\section{Spin 1/2: fermionic quadruplets}

Consider irreps of $SL(2,C)_{\mathrm{int}}$, whose weight diagrams are
determined by eigenvalues of generators of the right projections of spin
$\hat{S}_{3}^{R}$ and $\hat{B}_{3}^{R}$.

Any finite-dimensional irrep occurs in the decomposition of left (or right)
GRR with the multiplicity equal to its dimension. For instance, in the case
of $SL(2,C)$ the irrep $T_{[1/2\;0]}$ occurs in the decomposition of the
left GRR twice: columns $(z^1\; z^2)$ and $(\uz^1\; \uz^2)$ carry this representation;
analogously, the irrep $T_{[1/2\;0]}$ also occurs in the decomposition of
the right GRR twice (rows $(z^1\; \uz^1)$ and $(z^2\; \uz^2)$).

Linear functions of coordinates $z$ on the Lorentz group describe spin
$1/2$ particles. There are $4$ linearly independent functions that are
transformed differently under a change of a b.r.f.. These are
$z,\uz,\cc z,\cc\uz$ (we have omitted the usual \textquotedblleft
laboratory\textquotedblright\ indices, since $z^{1}$ and $z^{2}$ are
transformed in the same way under the action of $SL(2,C)_{\mathrm{int}}$).
These functions can be arranged into linear combinations being
eigenfunctions of various operators. The states of spin $1/2$ with a
definite charge $S_{3}^{R}$ and a chirality correspond to the functions
\begin{equation}
\begin{array}{lcc}
        &   S_3^R=-1/2 &  S_3^R=1/2 \\
R\;(\Gamma^5=1/2): \quad & e^{ipx}z^\alpha \quad   & e^{ipx}\uz^\alpha  \\
L\;(\Gamma^5=-1/2):\quad & e^{ipx}\cc\uz_\da \quad & e^{ipx}\cc z_\da   \\
&  &
\end{array}
\label{LRm}
\end{equation}
The states with a definite parity $\eta $ are eigenfunctions of the operator
$\hat P$, $\hat Pf(x,{z})=\eta f(x,{z})$; in the rest frame at $p_{0}=\pm m,$ we have
\begin{equation}
\begin{array}{lcc}
        &   S_3^R=-1/2 &  S_3^R=1/2 \\
\eta=1 \quad  & e^{\pm imx^0}(z^\alpha - \cc \uz_\da) \quad & e^{\pm imx^0}(\uz^\alpha + \cc z_\da) \\
\eta=-1\quad  & e^{\pm imx^0}(z^\alpha + \cc \uz_\da) \quad & e^{\pm imx^0}(\uz^\alpha - \cc z_\da) \\
&  &
\end{array}
\label{etam}
\end{equation}
We note that functions from (\ref{etam}) are eigenfunctions for $\hat{\Gamma}^{00}$
and $\hat{p}_{0}$, and, therefore, they are solutions of the left-invariant equation (\ref{fseq}).

Assuming a symmetry violation under right transformations of the Poincar\'{e}
group, and retaining only the particle-antiparticle symmetry, we find that a
mixed picture becomes possible:
\begin{equation}
\begin{array}{lcc}
       &   S_3^R=-1/2 &  S_3^R=1/2 \\
\quad  & e^{\pm imx^0}(z^\alpha - \cc \uz_\da) \quad & e^{ipx}\cc z_\da \\
\quad  & e^{ipx} z^\alpha                     \quad & e^{\pm imx^0}(\uz^\alpha - \cc z_\da) \\
&  &
\end{array}
\label{etaLRm}
\end{equation}

States from (\ref{LRm}) correspond to the massless or ultra-relativistic case,
when one has two chiral left fermions (for example, $e_{L}^{-}$ and $\nu _{L}$)
and two chiral right antifermions ($e_{R}^{+}$ and $\tilde{\nu}_{R}$), see Fig.1a.

The massive electron and neutrino correspond to states from (\ref{etam}).
In this case, the electron $e^{-}$ and the neutrino $\nu $ have a positive
internal parity; the positron $e^{+}$ and the antineutrino $\tilde{\nu}$
have a negative internal parity, see Fig.1b.

Finally, states from (\ref{etaLRm}) can be considered as a good
approximation within the range of energies much larger than the neutrino
mass. In this approximation, the quadruplet of leptons
$e^{-},\,\tilde{e}^{+},\,\nu _{L},\,\tilde{\nu}_{R}\,$ contains an electron with
a positive internal parity, a positron with a negative internal parity; a massless
neutrino can be only a left one (chirality is negative) and the antineutrino
can be only a right one (chirality is positive). In addition, a change of
signs both of $S_{3}^{R}$ and of chirality or parity corresponds to a
transition to an antiparticle, whereas a change of sign alone leads to a
transition to a state which is not present in (\ref{etaLRm}).

The quark quadruplet $u,\tilde{u},d,\tilde{d}$, where $u$ and $d$ are
characterized by the same internal parity $\eta =1$, and, according to
(\ref{s3rzar2}), by opposite signs of $S_{3}^{R}$, corresponds to (\ref{etam}).
In addition, as should be expected, particles and corresponding
antiparticles are characterized by opposite parity. However, in this case we
also encounter a violation of the symmetry with respect to right
transformations of the Poincar\'{e} group -- components of
$SL(2,C)_{\mathrm{int}}$-dublet have different mass.

This classification can be visualized in the form of a weight diagram of the
representation $T_{[1/2\; 0]}\oplus T_{[0\; 1/2]}$ of the group
$SL(2,C)_{\mathrm{int}}$.

\newsavebox{\spinhalfa}
\savebox{\spinhalfa}(160,180)[lb]
{
\put(0,70){\vector (1,0){150}}
\put(75,0){\vector (0,1){140}}
%\put(-2,143){$T_{[1/2\; 0]}\oplus T_{[0\; 1/2]}$}
\put(105,40){\circle*{4}}
\put(108,30){$\uz-\cc z,\,e^+$}
\put(45,100){\circle*{4}}
\put(0,90){$z-\cc \uz,\;e^-$}
\put(45,40){\circle*{4}}
\put(0,29){$z+\cc \uz,\;\tilde \nu$}
\put(105,100){\circle*{4}}
\put(108,89){$\uz+\cc z,\;\nu$}
\put(148,60){$S_3^R$}
\put(80,140){$\eta$}
\put(0,0){$b$}
\multiput(45,40)(5,0){12}{\circle*{1}}
\multiput(45,100)(5,0){12}{\circle*{1}}
}

\begin{figure}[ht]
{\caption{The weight diagram of the representation
$T_{[1/2\; 0]}\oplus T_{[0\; 1/2]}$ of $SL(2,C)_{\mathrm{int}}$.
The dotted line joins states related by transformations
of $SL(2,C)_{\mathrm{int}}$.
a) States with definite chirality, functions (\ref{LRm}).
b) States with definite parity $\eta$, functions (\ref{etam}).}
}
\begin{picture}(380,150)
\put(0,70){\vector (1,0){150}}
\put(75,0){\vector (0,1){140}}
%\put(-2,143){$T_{[1/2\; 0]}\oplus T_{[0\; 1/2]}$}
\put(105,40){\circle*{4}}
\put(108,30){$\cc z,\,\nu_L$}
\put(45,100){\circle*{4}}
\put(18,90){$\cc \uz,\,e^-_L$}
\put(45,40){\circle*{4}}
\put(18,29){$z,\,\tilde\nu_R$}
\put(105,100){\circle*{4}}
\put(108,89){$\uz,\,e^+_R$}
\put(148,60){$S_3^R$}
\put(80,140){$-iB_3^R$}
\put(0,0){$a$}
\put(200,0){\usebox{\spinhalfa}}
\multiput(45,100)(5,-5){12}{\circle*{1}} %{\line*(1,-1){5}}
\multiput(45,40)(5,5){12}{\circle*{1}}
\end{picture}
\end{figure}
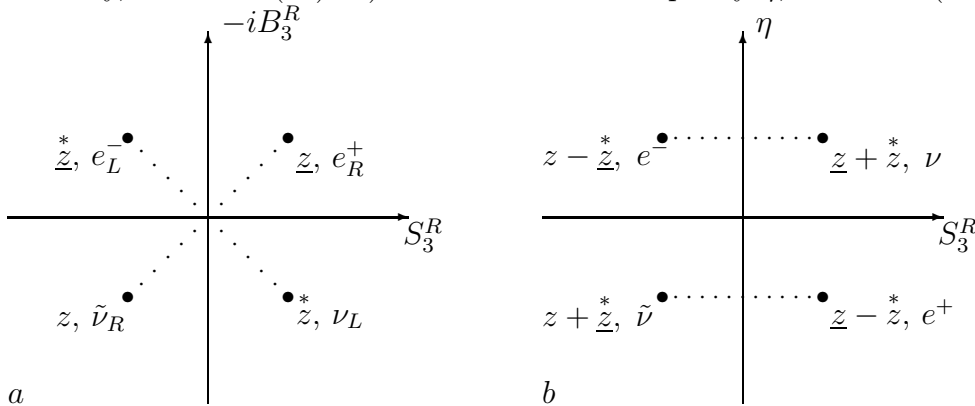

Note that besides the eigenfunctions of $\hat S_3^R$ one can also construct
states with definite charge parity, $\hat C f(x,{z})=\eta_c f(x,{z})$ (which
describe the Majorana neutrino), or with $\hat C\hat P\hat T$-parity (the so-called
``physical Majorana neutrino'', see \cite{KayGo83,KayGiP89}).

Let us now consider functions corresponding to a massive particle, moving
along the axis $x^{3}$. They can be obtained from functions in the rest
frame (\ref{etam}), which are characterized by a certain internal parity,
with the help of a Lorentz transformation
\begin{equation*}
P=UP_0U^\dagger, \quad Z=UZ_0, \quad \hbox{where}\;\; P_0=\pm\diag\{m,m\},
     \quad U=\diag\{e^{a},e^{-a}\}\in SL(2,C)_{\mathrm{ext}},
\end{equation*}
the sign of $P_{0}$ corresponds to the sign of $p_{0}$,
\begin{equation}
p_\mu=k_\mu{\rm sign}\ p_0, \quad k_0= m \cosh 2a, \quad k_3= m \sinh 2a,
\quad  e^{\pm a}=\sqrt{(k_0\pm k_3)/m}.
\label{sol2a}
\end{equation}
By applying these transformations to the state with $S_{3}^{R}=-1/2$, $\eta
=1$ at $p_{0}>0$, we find
\begin{equation*}
f'_{m,1/2}(x,z)=  e^{i(k_0x^0+k_3x^3)}\left[
 C_1(z^1e^a-\cc\uz_{\dot 1}e^{-a}) + C_2(z^2e^{-a}-\cc\uz_{\dot 2}e^a)\right],
\end{equation*}
where the first term in the square brackets corresponds to $s_{3}=1/2$, and
the second term corresponds to $s_{3}=-1/2$. In the ultra-relativistic case
with a positive $a$ (i.e. with $k_{3}>0$) there remain only two components,
\begin{equation*}
f'_{m,1/2}(x,\bz) \approx   e^{i(k_0x^0+k_3x^3)}\left(
 C_1 z^1e^a - C_2\cc\uz_{\dot 2}e^a\right).
\end{equation*}
which are eigenfunctions of the helicity operator $\hat{p}\hat{S}$ with the
eigenvalues $p_{3}s_{3}=\frac{1}{2}k_{3}\mathop{{\rm sign}}s_{3}$; these
components are also eigenfunctions of the operator $\hat{\Gamma}^{5}$ with
the same sign. In a similar way, considering the case $a<0$ and other states
from (\ref{etam}), we conclude that in the ultra-relativistic limit with
$p_{0}>0$ signs of chirality $\hat{\Gamma}^{5}$ and helicity $\hat{p}\hat{S}$
are the same.
We stress, that the above conclusions derived for the ultra-relativistic
case coincide with the results obtained from the Dirac equation.

Consider now the states corresponding to spin $1/2$ particles from the
viewpoint of solutions of left-invariant RWE of first order in more detail.
Equation (\ref{fseq}) for functions linear in $z$ splits into a pair of
Dirac equations for functions from subspaces $f(x,z,\cc\uz)$ ($S_3^R=-1/2$)
and $f(x,\uz, \cc z)$ ($S_3^R=1/2$). The sign of the mass term in these
equations is $\varepsilon =\eta \sign p_{0} \sign S_{3}^{R}$
(see \cite{GitSh09,GitSh01,BucGiS02} for details).

For spin $s=1/2$, eigenfunctions of the operator $\hat{S}_{3}^{R}$ and space
parity $\hat P=\hat I_{s}$ are
$z^\alpha \pm \cc\uz_{\da}$, $\uz^\alpha \pm \cc z_{\da}$.
In the rest frame solutions of two mentioned above Dirac equations
with $\varepsilon =1$ have the form
\begin{eqnarray}
&&f_1(x,z)=e^{imx^0}C_\alpha(z^\alpha-\cc\uz_{\da})+e^{-imx^0} C'_\alpha(z^\alpha+\cc\uz_{\da}), \quad S_3^R=-1/2,
\label{DirSol1}
\\
&&f_2(x,z)=e^{imx^0}D_\alpha(\uz^\alpha-\cc z_{\da})+e^{-imx^0} D'_\alpha(\uz^\alpha+\cc z_{\da}), \quad S_3^R=1/2.
\label{DirSol2}
\end{eqnarray}
As is known, the free Dirac equation have solutions corresponding to a pair
of non-equivalent irreps of the improper Poincar\'{e} group with opposite
signs of $\eta $ and $p_{0}$. Consider the solution (\ref{DirSol1}) of the
first equation. Assuming, as usual, that the wave-function of an
antiparticle is a bispinor, being charge-conjugated to a certain
negative-frequency solution of the Dirac equation \cite{BerLiP}, we find that
the antiparticle is associated with the function
$e^{imx^0}\cc C'_\alpha (-1)^{\alpha}(\uz^\alpha-\cc z_{\da})$, being
complex-conjugated to the negative-frequency part of solution (\ref{DirSol1})
($\hat C\hat P\hat T$-conjugation yields $e^{imx^0}C'_\alpha (\uz^\alpha-\cc z_{\da}$).
This function is a solution of the equation (\ref{fseq})
with the same $\varepsilon =1$, see the first term of (\ref{DirSol2});
however, it is characterized by opposite signs of $\eta $ and $S_{3}^{R}$ ($\eta=-1$, $S_{3}^{R}=1/2$).
Thus, a particle (electron) and a antiparticle (positron)
are described by positive-frequency solutions of eq. (\ref{fseq}) with
$\varepsilon =1$ and opposite signs of $S_{3}^{R}$ and $\eta $.

There remain two unused functions,
$e^{imx^0}(z^\alpha+\cc\uz_{\da})$ and $e^{imx^0}(\uz^\alpha+\cc z_{\da})$,
that correspond to a particle with a negative parity and an antiparticle
with a positive parity, which provides solutions of the Dirac equations with
$\varepsilon =-1$,
\begin{eqnarray}
&&f_3(x,z)=e^{imx^0}C_\alpha(z^\alpha+\cc\uz_{\da})+e^{-imx^0} C'_\alpha(z^\alpha-\cc\uz_{\da}), \quad S_3^R=-1/2,
\label{DirSol3}
\\
&&f_4(x,z)=e^{imx^0}D_\alpha(\uz^\alpha+\cc z_{\da})+e^{-imx^0} D'_\alpha(\uz^\alpha-\cc z_{\da}), \quad S_3^R=1/2.
\label{DirSol4}
\end{eqnarray}
However, these functions with $p_{0}>0$ can't describe electron or positron:
does not exist an electron with negative parity and a positron with positive
parity.

Consequently, it is natural to associate the remaining functions with
another particles, as was done above.

\section {Spin 1}

Spin $1$ particles are described by quadratic combinations of ${z}$, which
are transformed with respect to the representations of the Lorentz group
$T_{[j_{1}j_{2}]}$, $j_{1}+j_{2}=1$. These are a $6$-dimensional
adjoint representation $T_{[10]}\oplus T_{[01]}$ (two complex-conjugate
matrices from $SO(3,C)$) and a $4$-dimensional vector representation
$T_{[\frac{1}{2}\frac{1}{2}]}$ (matrix from $SO(3,1)$).

One can see that if the spin part of a wave function is transformed
according to the representation $T_{[s0]}$ or $T_{[0s]}$ then the eigenvalue
of the Casimir operator $W^{2}$ is equal to $-m^{2}s(s+1)$, i.e., $s$ is
spin. Therefore, all the states carrying the representation
$T_{[10]}\oplus T_{[01]}$ have spin $1$.

For this representation all $6$ states are characterized by chirality
$\Gamma ^{5}=j_{1}-j_{2}=\pm 1$. The multiple weight $S_{3}^{R}=B_{3}^{R}=0$,
being in the center of the diagram (Fig.2a), corresponds to a pair of
states. Notice that states with a definite parity $V_{\eta =\pm 1}^{0}$,
corresponding to $S_{3}^{R}=0$ (Fig.2b), are also characterized by a
definite charge parity, and therefore they can describe real neutral particles.

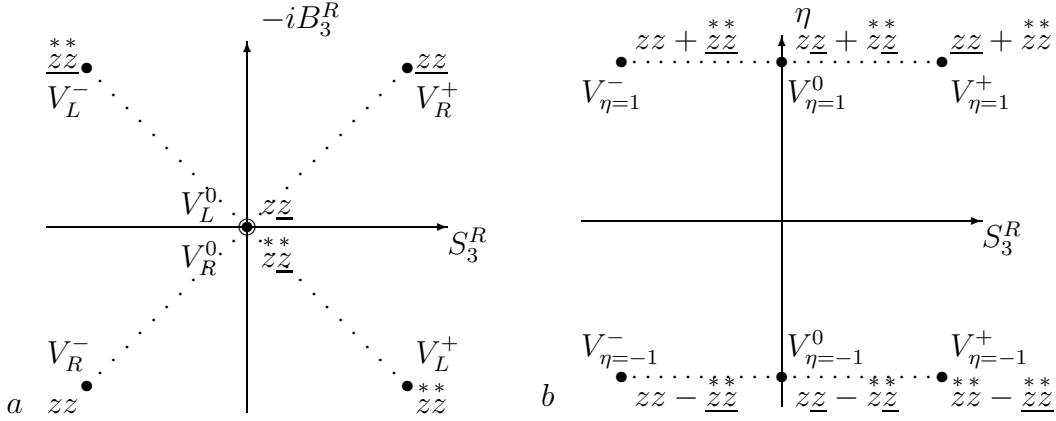
\begin{figure}[ht]
\caption{The weight diagrams of the representation $T_{[1\; 0]}\oplus T_{[0\; 1]}$ of
$SL(2,C)_{\mathrm{int}}$, $\Gamma^5=\pm 1$.
(a), $\protect\eta=\pm 1$
(b). The dotted lines join states related by the transformations of $SL(2,C)_{\mathrm{int}}$.}

\newsavebox{\spinonea}
\savebox{\spinonea}(160,165)[lb]
{
\put(0,70){\vector (1,0){150}}
\put(75,0){\vector (0,1){140}}
%\put(0,153){$T_{[1\; 0]}\oplus T_{[0\; 1]}$}
\put(20,135){$zz+\cc \uz \cc \uz$}
\put(0,115){$V^-_{\eta=1}$}
\put(138,0){$\cc z \cc z-\cc\uz\cc\uz$}
\put(15,11){\circle*{4}}
\put(135,11){\circle*{4}}
\put(15,130){\circle*{4}}
\put(135,130){\circle*{4}}
\put(20,0){$zz-\cc\uz\cc\uz$}
\put(0,19){$V^-_{\eta=-1}$}
\put(138,135){$\uz\uz+\cc z\cc z$}
\put(138,19){$V^+_{\eta=-1}$}
\put(138,115){$V^+_{\eta=1}$}
\put(75,11){\circle*{4}}
\put(75,130){\circle*{4}}
\put(80,135){$z\uz+\cc z\cc \uz$}
\put(80,0){$z\uz-\cc z\cc \uz$}
\put(77,19){$V^0_{\eta=-1}$}
\put(77,115){$V^0_{\eta=1}$}
\put(150,60){$S_3^R$}
\put(80,145){$\eta$}
\put(-15,0){$b$}
\multiput(15,130)(5,0){25}{\circle*{1}}
\multiput(15,11)(5,0){25}{\circle*{1}}
}
\par
\begin{picture}(360,165)
\put(0,70){\vector (1,0){150}}
\put(75,0){\vector (0,1){140}}
%\put(0,153){$T_{[1\; 0]}\oplus T_{[0\; 1]}$}
\put(0,130){$\cc \uz \cc \uz$}
\put(0,115){$V^{-}_L$}
\put(138,0){$\cc z \cc z$}
\put(15,10){\circle*{4}}
\put(135,10){\circle*{4}}
\put(15,130){\circle*{4}}
\put(135,130){\circle*{4}}
\put(0,0){$zz$}
\put(0,19){$V^{-}_R$}
\put(138,130){$\uz\uz$}
\put(138,19){$V^{+}_L$}
\put(138,115){$V^{+}_R$}
\put(75,70){\circle*{4}}
\put(75,70){\circle{6}}
\put(80,75){$z\uz$}
\put(80,55){$\cc z\cc \uz$}
\put(50,75){$V^0_L$}
\put(50,55){$V^0_R$}
\put(150,60){$S_3^R$}
\put(80,145){$-iB_3^R$}
\put(-15,0){$a$}
\multiput(15,130)(5,-5){25}{\circle*{1}}
\multiput(15,10)(5,5){25}{\circle*{1}}

\put(200,0){\usebox{\spinonea}}
\end{picture}
\end{figure}

By restoring laboratory indices, one can easily see that each point of the
weight diagram (Fig.2) corresponds to three states (according to the number
of possible spin projections), that transform equally under
$SL(2,C)_{\mathrm{int}}$, but differently under $SL(2,C)_{\mathrm{ext}}$. In
particular, for states with $S_{3}^{R}=0$ among four pairwise products three
functions correspond to spin $1$, namely
\begin{equation*}
z^1\uz^1,\; z^2\uz^2,\; z^1\uz^2+z^2\uz^1,
\end{equation*}
since, due to unimodularity, $\det Z = z^1\uz^2-z^2\uz^1=1$ is a Lorentz scalar.

Making a reduction to the compact group $SU(2)$, we obtain two triplets:
left and right, corresponding to the diagonals on Fig.2a, or triplets with a
fixed parity, Fig.2b.

Consider now the representation $T_{[\frac{1}{2}\frac{1}{2}]}$ (Fig.3).
\begin{figure}[th]
\caption{The weight diagrams of the representation $T_{[\frac{1}{2}\;\frac{1}{2}]}$
of $SL(2,C)_{\mathrm{int}}$, $\Gamma ^{5}=0$.\quad
On Fig.a the dotted line joins states related by the transformations of
$SL(2,C)_{\mathrm{int}}$,
On Fig.b the dotted line joins states related by the transformations of $SU(2)\subset SL(2,C)_{\mathrm{int}}$.}
\newsavebox{\spinoneb}
\savebox{\spinoneb}(160,160)[lb]
{
\put(0,70){\vector (1,0){150}}
\put(75,0){\vector (0,1){140}}
%\put(0,153){$T_{[\qq\;\qq]}$}
\put(7,115){$z \cc \uz$}
\put(127,115){$\uz\cc z$}
\put(78,0){$z \cc z+\uz \cc \uz$}
\put(78,115){$z \cc z-\uz \cc \uz$}
\put(75,10){\circle*{4}}
\put(75,130){\circle*{4}}
\put(15,130){\circle*{4}}
\put(135,130){\circle*{4}}
\put(150,60){$S_3^R$}
\put(80,140){$\eta$}
\put(143,125){$S^R\!=1$}
\put(143,5){$S^R\!=0$}
\put(0,0){$b$}
\multiput(15,130)(5,0){25}{\circle*{1}}
}

\begin{picture}(360,160)
\put(0,70){\vector (1,0){150}}
\put(75,0){\vector (0,1){140}}
%\put(0,153){$T_{[\qq\;\qq]}$}
\put(7,53){$z \cc \uz$}
\put(127,53){$\uz\cc z$}
\put(78,0){$z \cc z$}
\put(78,120){$\uz \cc \uz$}
\put(75,10){\circle*{4}}
\put(75,130){\circle*{4}}
\put(15,70){\circle*{4}}
\put(135,70){\circle*{4}}
\put(150,60){$S_3^R$}
\put(80,140){$-iB_3^R$}
\put(0,0){$a$}
\multiput(15,70)(5,-5){12}{\circle*{1}}
\multiput(15,70)(5,5){12}{\circle*{1}}
\multiput(135,70)(-5,-5){12}{\circle*{1}}
\multiput(135,70)(-5,5){12}{\circle*{1}}
\put(200,0){\usebox{\spinoneb}}
\end{picture}
\end{figure}
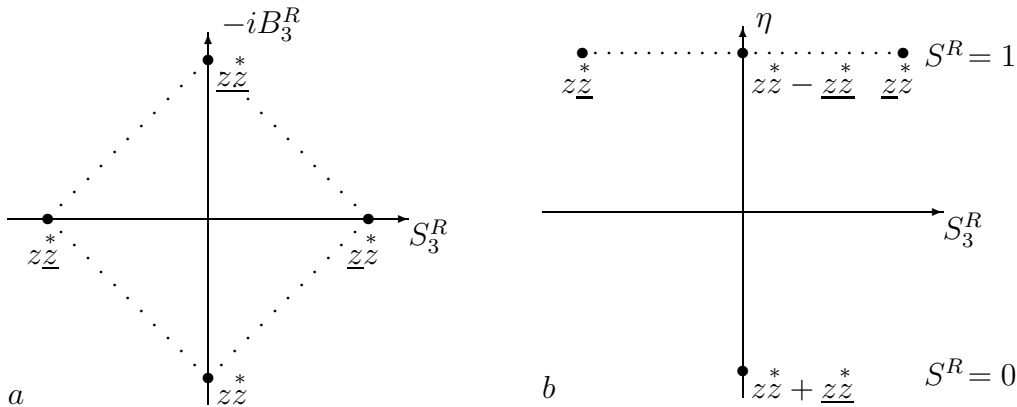
For this representation, all states have the chirality $\Gamma ^{5}=0$. The
reduction to the compact subgroup $SU(2)$ gives a triplet and singlet, which
can be seen on Fig.3b. Besides, in contrast to $T_{[1\;0]}\oplus T_{[0\;1]}$,
for $T_{[\frac{1}{2}\frac{1}{2}]}$ possible values of spin are $1$ and $0$.
Thus, to describe spin $1$-particles, functions $f(x,{z})$, carrying
representation $T_{[\frac{1}{2}\frac{1}{2}]}$, must satisfy certain
subsidiary conditions. In particular, in the subspace of functions
$f(x,z,\cc\uz)$ at $\hat{\mathrm{p}}^{2}=m^{2}$ we have \cite{GitSh01}
\begin{equation}
\hat{W}^{2}=-m^{2}(j_{1}+j_{2})(j_{1}+j_{2}+1)+4\hat{p}_{\mu }q^{\mu }\hat{p}_{\nu }\hat{V}^{\nu },
\label{31W}
\end{equation}
where
$q^\mu =\frac 12  \sigma^\mu_{\;\;\alpha\da} z^{\alpha}\cc\uz^{\dot\alpha}$ and
$\hat V^\mu=\frac 12  \sigma^\mu_{\;\;\alpha\da} \partial^{\alpha}\upartial^{\da}$.
Consequently, in the case $s=j_{1}+j_{2}$ a necessary and sufficient condition of spinorial
irreducibility is given by
\begin{equation}
\hat{p}_{\mu }q^{\mu }\hat{p}_{\nu }\hat{V}^{\nu }f(x,z,\cc\uz)=0.
\label{sub0}
\end{equation}
For the representations $T_{[s\,0]}$ and $T_{[0\,s]}$, this condition is
fulfilled identically, since in this case $\hat{V}^{\mu }f(x,z)=0$. In the
general case, taking into account that in the momentum representation, the
action of the operator $q^{\mu }\hat{p}_{\mu }$ is reduced to multiplication
by a number, we arrive at alternative conditions,
\begin{eqnarray}
&&p_{\mu }q^{\mu }=0,  \label{V1p} \\
&&\hat{p}_{\nu }\hat{V}^{\nu }f(x,z,\cc\uz)=0.  \label{V2p}
\end{eqnarray}
In the first case, we have a space of functions of two $4$-vectors
$p_{\mu }$, $q_{\mu }$, which are subject to invariant constraints,
\begin{equation}
p^{2}=m^{2},\quad p_{\mu }q^{\mu }=0,\quad q^{2}=0.  \label{pqconstr}
\end{equation}
In the rest frame, according to (\ref{pqconstr}), we have
$z^{1}\cc\uz^{1}+z^{2}\cc\uz^{2}=0$. Such an approach to the construction of wave
functions, describing elementary particles, was suggested by Wigner in the
article \cite{Wig63}, where the discussion was restricted to particles of
integer spin and to real-valued $q_{\mu }$ with the constraints $p^{2}=m^{2}$,
$p_{\mu }q^{\mu }=0$, $q^{2}=-1$. Different generalizations of the
Wigner's approach were considered in \cite{KimWi87,BieBrT88,HasSi92,KuzLy95,LyaSeS96}.

The second requirement (\ref{V2p}) is only a condition on the functions
$\phi (x)$ and it does not concern the spinorial variables. Indeed,
representing the function $f(x,{z})$ in the form
\begin{equation}
f(x,z)= \phi_{\alpha}^{\;\;\db}(x) z^{\alpha} \cc\uz_{\dot\beta}=\Phi_\mu (x) q^\mu , \qquad
\Phi_\mu(x)= -\bar\sigma_\mu^{\;\;\db\alpha} \phi_{\alpha\db}(x), \quad
\end{equation}
we find \cite{GitSh01} that $\Phi _{\mu }(x)$ obey equations for spin
$1$-particles in the Proca form
\begin{equation}
(\hat{p}^{2}-m^{2})\Phi _{\mu }(x)=0,\qquad \hat{p}^{\mu }\Phi _{\mu }(x)=0.
\label{Proca}
\end{equation}

Above we, considering possible values of $S_{3}^{R}$ for $W^{\pm }$ bosons,
have excluded $S_{3}^{R}=0$. Here we see that all states of spin
$1$ with $j_{1}+j_{2}=1$ and $S_{3}^{R}=0$ are related to
true neutral particles and can't be related with $W^{\pm }$.

Therefore, in the framework of the present theory, we arrive to two families
of particles of spin $1$: two triplets, corresponding to $T_{[1\;0]}\oplus T_{[0\;1]}$
and a quadruplet corresponding to $T_{[\frac{1}{2}\;\frac{1}{2}]}$,
which (since a reduction to a compact subgroup was done) is decomposed
into a triplet with $\eta =1$ and a singlet with $\eta =-1$. They
are candidates to describe well-known particles of spin $1$ -- the triplet
of intermediate vector bosons and the photon.

To give an exact answer, let us consider the weight diagram of
$6$-dimensional adjoint irrep of the Lorentz group $T_{[1\;0]}\oplus T_{[0\;1]}$
(Fig.2). The multiple weight $S_{3}^{R}=0$ can be related with real neutral
particles -- the photon $\gamma $ and the $Z^{0}$-boson, the weights
with $S_{3}^{R}=1$ and $S_{3}^{R}=-1$ can be related with $W^{+}$ and $W^{-}$
bosons. In addition, each of the latter appears twice as $W_{L}^{\pm }$ and
$W_{R}^{\pm }$ (linear combinations correspond to states with a definite
parity $W_{\eta =\pm 1}^{\pm }$).

As far as the $4$-dimensional representation $T_{[\frac{1}{2}\;\frac{1}{2}]}$
is concerned, the $W^{+}$ and $W^{-}$ bosons can be associated with states
having $S_{3}^{R}=\pm 1$, with the parity $\eta =1$, whereas the photon and
$Z^{0}$-boson can be associated with two states having the zero charge
$S_{3}^{R}$, see Fig.3b.

However, more detail consideration exclude the case $T_{[1\;0]}\oplus T_{[0\;1]}$.
In massless limit not only $S_{3}^{R}$, but also $B_{3}^{R}$
and $\Gamma ^{5}$ are conserved quantum numbers. Then, $e_{L}^{-}$ and
$\tilde{\nu}_{R}$ are characterized by $\Gamma ^{5}=1/2$ and $\Gamma ^{5}=-1/2$
(see Fig.1 and (\ref{LRm})), so for $W^{-}$ we have $\Gamma ^{5}=0$. The
latter is fulfilled for $T_{[\frac{1}{2}\;\frac{1}{2}]}$, but not for
$T_{[1\;0]}\oplus T_{[0\;1]}$. Analogously, it is easy to see (Fig.1) that
$e_{L}^{-}$ and $\tilde{\nu}_{R}$ are characterized by opposite values of
$B_{3}^{R}$, and therefore the charged $W^{-}$ ($S_{3}^{R}=-1$) must have
$B_{3}^{R}=0$, which holds true for states with $S_{3}^{R}=-1$, described by
the representation $T_{[\frac{1}{2}\;\frac{1}{2}]}$ (Fig.3), but not by
$T_{[1\;0]}\oplus T_{[0\;1]}$ (Fig.2).

\section{Quasiregular representations and spin description \newline
(geometrical models of spinning particles)}

The consideration of GRR of the Poincar\'e group ensures the possibility of
consistent description of particles with arbitrary spin by means of scalar
functions on $\mathcal{M}\times \mathrm{Spin}(3,1)$, where $\mathcal{M}$ is
Minkowski space. At the same time, for description of spinning particles it
is possible to use the spaces $\mathcal{M}\times L$, where $L$ is some
homogeneous space of the Lorentz group (one or two-sheeted hyperboloid,
cone, projective space and so on); see, for example, \cite%
{BacKi69,Kihlb70,BoyFl74,Wig63,KimWi87,BieBrT88,HasSi92,KuzLy95,LyaSeS96,DerGi99,Drech97,Varla05}.
In some papers fields on homogeneous spaces are considered; in other
papers such spaces are treated as phase spaces of some classic mechanics,
and the latter are treated as models of spinning relativistic particles.

These spaces appear in the framework of the next group-theoretical scheme.
Let us consider the left quasiregular representation of the Poincar\'{e} group
\begin{equation}
T(g)f(g_{0}K)=f(g^{-1}g_{0}K),\quad K\subset \mathrm{Spin}(3,1),
\label{kvazi.1}
\end{equation}
and since $x$ is invariant under right rotations (see (\ref{gen.02}))
\begin{equation*}
g_{0}\leftrightarrow (X,Z),\quad g_{0}K\leftrightarrow (X,ZK).
\end{equation*}
Therefore the relation (\ref{kvazi.1}) defines the representation of the
Poincar\'{e} group in the space of functions $f(x,zK)$ on
\begin{equation}
\mathcal{M}\times (\mathrm{Spin}(3,1)/K).
\end{equation}
Generally speaking, in the space of scalar functions on $\mathrm{Spin}(3,1)/K $
one can realize only a part of irreps of the Lorentz group, and in
the space of scalar functions on $\mathcal{M}\times (\mathrm{Spin}(3,1)/K)$
one can realize only a part of irreps of the Poincar\'{e} group. In
particular, the case $K=\mathrm{Spin}(3,1)$ corresponds to a scalar field on
Minkowski space.

Thus the consideration of left quasiregular representations allows one to
construct a number of spin models classified by subgroups
$K\subseteq \mathrm{Spin}(3,1)$. However the dimension of the space $M(3,1)/K$ is
reduced in comparison with $M(3,1)$ by the number of generators of the group
$K$, respectively the number of commuting operators and the number of
quantum numbers is reduced as well.

There exist $13$ homogeneous spaces $M(3,1)/K$, containing Minkowski space
\cite{Finke55,BacKi69,Kihlb70}. We will consider $5$ such spaces which were
used in constructing different geometrical models.
\bigskip

{Some homogeneous spaces related to the group $SL(2,C).$}

\noindent
\begin{tabular}{|l|p{1.61in}|l|l|p{1.1in}|l|}
\hline
\phantom{1}
& \multicolumn{2}{|c|}{\begin{tabular}{l} Name and dimension \\ of space \end{tabular}}
& \begin{tabular}{l} Elements and \\ Transformations \end{tabular}
& \begin{tabular}{l} Subgroup $K_i$ \end{tabular}
& \begin{tabular}{l} Internal \\ numbers \end{tabular} \\
\hline
1
& Complex affine plane
& 4
& $\begin{array}{c} (z^1,z^2)\rightarrow  \\ (\alpha z^1\!+\!\gamma z^2,\beta z^1\!+\!\delta z^2) \end{array} $
& $\left( \begin{array}{cc} 1 & \zeta  \\ 0 & 1 \end{array} \right) $
& $j_1, j_2$
\\
\hline
2
&\vspace{-0.4cm}{\raggedright Complex projective \\ line $\mathbb C P^1\sim S^2$ }
& 2
& \begin{tabular}[c]{l}$z=\displaystyle\frac{z^1}{z^2}$, \quad $z\rightarrow \frac{\alpha z+\gamma }{\beta z+\delta }$\end{tabular}
& $\left( \begin{array}{cc} \alpha  & \beta  \\ 0 & \alpha ^{-1} \end{array} \right) $
& --
\\
\hline
3
& \vspace{-0.6cm}{\raggedright Lobachevskian 3-space \\ (positive sheet of \\ $H_{1}^{1,3}$ hyperboloid)}
& 3
& $\begin{array}{c} Q\rightarrow UQU^{\dagger }, \\ \det Q=1 \end{array} $
& \begin{tabular}{l} $\left( \begin{array}{cc} \alpha  & \beta  \\ -\bar \beta  & \bar \alpha \end{array} \right) ,$
  \\ $|\alpha |^2+|\beta |^2=1$ \end{tabular}
& $j_1=j_2,\; \eta$
\\
\hline
4
&\vspace{-0.6cm}{\raggedright Imaginary \\ Lobachevskian 3-space \\ ($H_{-1}^{1,3}$ hyperboloid) }
& 3
& $\begin{array}{c} Q\rightarrow UQU^{\dagger }, \\ \det Q=-1 \end{array} $
& \begin{tabular}{l} $\left( \begin{array}{cc} \alpha  & \beta  \\ \bar \beta  & \bar \alpha \end{array} \right) ,$
  \\$|\alpha |^2-|\beta |^2=1$\end{tabular}
& $j_1=j_2,\; \eta$
\\
\hline
5
& {\raggedright The cone $H_0^{1,3}$}
& 3
& $\begin{array}{c} Q\rightarrow UQU^{\dagger }, \\ \det Q=0 \end{array} $
& $\left( \begin{array}{cc}e^{-i\varphi } & 0 \\ \zeta  & e^{i\varphi } \end{array} \right) $
& $j_1=j_2$
\\
\hline
\end{tabular}
\bigskip

Here $z^{1},z^{2}$, and ${\underline{z}}{\vphantom{z}}^{1},{\underline{z}}{%
\vphantom{z}}^{2}$ are elements of the first and the second columns of the
matrix $Z\in SL(2,C)$, $2\times 2$ matrix $Q$ corresponds to $4$-vector $%
q^{\mu }$,
\begin{equation*}
Q=\left( \begin{array}{cc} q^0+q^3 & q^1-iq^2 \\ q^1+iq^2 & q^0-q^3
\end{array} \right), \qquad
U=\left( \begin{array}{cc} \alpha & \beta \\ \gamma & \delta \end{array} \right),
\qquad \det U=1.
\end{equation*}
The latter vectors have different expressions in terms of $z$ in different
cases. For a cone,
$q^\mu=\sigma^{\mu}_{\;\;\dot\beta\alpha} z^\alpha \cc z^{\dot\beta}$
\cite{GelGrV66}; in two other cases $q^{\mu }$ is expressed
via the tetrads
$v^\mu_{\;\;\nur}=\sigma^{\mu}_{\;\;\dot\beta\alpha} \sigma_{\nur}^{\;\;\alphar\dot\alphar} z^\beta_{\;\;\betar} \cc \uz^{\dot\beta}_{\;\;\dot\alphar}$
\cite{GitSh09}, then we have $q^{\mu }=v_{\;\;\underline{0}}^{\mu }$
for the subgroup $K_{4}=SU(2)$, different $SU(1,1)$-subgroups correspond to
$v_{\;\;\underline{1}}^{\mu },v_{\;\;\underline{2}}^{\mu },$
$v_{\;\;\underline{3}}^{\mu }$ or to their linear combinations.

Let us discuss quantum numbers that can label quantum states corresponding
to scalar functions
\begin{equation}
f'(y')=f(y), \quad f'(y)=T(g)f(y)=f(g^{-1})y,\; y'=gy, \quad y\in M(3,1)/K,
\label{kvazi.3}
\end{equation}
defined on the above listed spaces.

1. The scalar field $f(x^{\mu },z^{\alpha },\cc z_{\dot{\alpha}})$ on the
$8$-dimensional space $M(3,1)/K_{1}$ (spinning space $SL(2,C)/K_{1}$ -- a
complex affine plane) depends on elements $z^{\alpha}$ of the first column
of the matrix $Z$ and complex conjugates $\cc z_{\dot{\alpha}}$. Such a
field was studied in \cite{BacKi69,Kihlb70,Drech97}. It possesses four
characteristics related to the orientation variables $z$ (the spin, its
projection, and a pair $(j_{1},j_{2})$ that fixes irrep of the Lorentz
subgroup). To a given irrep of the Lorentz subgroup correspond homogeneous
functions of the power $2j_{1}$ in $z^{\alpha }$ and $2j_{2}$ in $\cc
z_{\dot{\alpha}}$. As follows from (\ref{SR}), eigenvalues of the generators
$\hat{S}_{R}^{3}$ and $\hat{B}_{R}^{3}$ are fixed, they are expressed in
terms of $j_{1}$ and $j_{2}$,
\begin{equation*}
S_{R}^{3}=j_{2}-j_{1},\;iB_{R}^{3}=(j_{2}+j_{1}).
\end{equation*}
In contrast to the case of functions on $M(3,1)$, the space $M(3,1)/K_{1}$
is not invariant under space reflection,
$Z\overset{\hat P}{\rightarrow }(Z^{\dagger })^{-1}$ or
$z^\alpha\to -\cc \uz_{\dot\alpha}$, $\cc \uz_{\dot\alpha}\to z^\alpha$
\cite{GitSh09}, and functions $f(x^{\mu },z^{\alpha },\cc z_{\dot{\alpha}})$
of elements of the first column convert to functions
$f(x^\mu,-\cc\uz_{\dot\alpha},\uz^\alpha)$ of elements of the second
column. Thus, states with a given parity cannot be described by scalar
functions on $M(3,1)/K_{1}$.

2. Let us consider a projective model. The $6$-dimensional space
$M(3,1)/K_{2}$ (the spinning space $SL(2,C)/K_{2}$ is a $2$-dimensional
sphere) is a space of the least dimensions which can provide a spin
description by one-component functions. Particle models in this space were
studied in detail in \cite{KuzLy95,LyaSeS96}. A relation to the previous
(spinor) model are given by the relations
\begin{equation*}
\phi (x^\mu, z,\cc z)= f(x^\mu, z,1,\cc z,1), \qquad
f(x^\mu, z^1,z^2,\cc z^1,\cc z^2)= (z^2)^{2j_1}(\cc z^2)^{2j_2}\phi (x^\mu, z,\cc z),
\end{equation*}
where\ $z={z^{1}}/{z^{2}}$. Here, $z$ are transformed linear fractionally in
contrast to other models where $z$ are transformed linearly. It is easily to
see that the transformation lows of same functions $\phi (x^{\mu },z,\bar{z})$
under the Lorentz group $SL(2,C)_{\mathrm{ext}}$ depend on $j_{1},j_{2}$
(the corresponding generators depend on $j_{1},j_{2}$),
\begin{equation}
\phi (x^\mu, z,\cc z) \to ({z^{2\prime}})^{2j_1}({\cc z^{2\prime}})^{2j_2}\phi ({x^\mu}', z',\cc z').
\label{kvazi.5}
\end{equation}
In particular, this means that the functions $\phi (x^{\mu },z,\cc z)$ do
not contain any information about a Lorentz group representation. We note
that transformations (\ref{kvazi.5}) of the functions $\phi (x^{\mu },z,\cc z)$
are not reduced to an argument change, and such functions are not scalar
ones with respect to the definition (\ref{kvazi.3}).

Space reflection transform $z$ into $\uz=\uz^1/\uz^2$, which
means, as in the previous case, that states with a given inner parity cannot
be described by functions $\phi (x^{\mu },z,\bar{z})$.

3. Vector models use functions $f(p_{\mu },q_{\mu })$ of $4$-momentum
$p_{\mu }$ and a spinning variable $q_{\mu }$,
\begin{equation}
\hat{S}_{\mu \nu }=i(q_{\mu }\partial {q^{\nu }}-q_{\nu }\partial {q^{\mu }}),\quad
\hat{\mathbf{S}}\hat{\mathbf{B}}=0.
\end{equation}
Since the Casimir operator $\hat{\mathbf{S}}\hat{\mathbf{B}}$ of the Lorentz
group is zero, we have $j_{1}=j_{2}$. A reduction of a irrep
$T_{[j_{1},j_{1}]}$ of the Lorentz group to a compact rotation subgroup is
given by the equation
\begin{equation}
\textstyle T_{[j_{1},j_{1}]}=\sum_{j=0}^{2j_{1}}T_{j}.  \label{qvazi.7}
\end{equation}
Thus, the models correspond to particles with integer spins that are
described by the representation $T_{[j_{1},j_{1}]}$ of the Lorentz group.

A $4$-vector\ $q_{\mu }$ is given by point on the hyperboloid $H_{-1}^{1,3}$,
$H_{1}^{1,3}$ or on the cone $H_{0}^{1,3}$ (see the table), respectively
$q_{\mu }q^{\mu }=0,\pm 1$. The condition\ $p_{\mu }q^{\mu }=0$ can be used
to select states with a spin $s=2j_{1}$ maximal for a given irrep
$T_{[j_{1},j_{1}]}$.

Thus, we have a family of models with scalar functions $f(p_{\mu },q_{\mu })$
and constraints
\begin{equation}
p_{\mu }p^{\mu }=m^{2},\qquad p_{\mu }q^{\mu }=0,\qquad q_{\mu }q^{\mu}=0,\pm 1.
\end{equation}

The spaces of functions $f(p_{\mu },q_{\mu })$ on the hyperboloids is
invariant under space reflection,
$q^{\mu }=v_{\;\;\underline{0}}^{\mu }\rightarrow -(-1)^{\delta _{\mu 0}}q^{\mu }$,
$q^{\mu }=v_{\;\;\underline{3}}^{\mu }\rightarrow (-1)^{\delta _{\mu 0}}q^{\mu }$,
and therefore, such spaces can serve to describe states with a definite inner parity $\eta $.
Functions on the cone depending on
$q^{\mu }=\sigma _{\;\;\dot{\beta}\alpha}^{\mu }z^{\alpha }\cc z^{\dot{\beta}}$
under space reflection are converted to functions of
$\underline q^\mu=\sigma^{\mu}_{\;\;\dot\beta\alpha} \uz^\alpha \cc \uz^{\dot\beta}$.
That is why one cannot construct such scalar functions corresponding to states with a
definite parity $\eta $.

Thus, scalar functions on the homogeneous spaces $\mathcal{M}\times (\mathrm{%
Spin}(3,1)/K)$, $K\subset \mathrm{Spin}(3,1)$, describe spinning particles,
however, they correspond to states where a part of inner (right) quantum
numbers is fixed (i.e., they are expressed via other quantum numbers) or are
not defined at all. In this case, the number of commuting operators and the
number of quantum numbers that characterize the field is reduced by the
number of generators of the subgroup $K$.

\section{Concluding remarks}

Orientable objects are described by the field $f(x,z)$ on the Poincar\'{e}
group. Functions $f(x,z)$ depend on $10$ parameters and admit two kinds of
transformations -- left (change of space-fixed reference frame, or Lorentz
transformations) and right (change of body-fixed reference frame). These
transformations form the direct product
$M(3,1)_{\mathrm{ext}}\times M(3,1)_{\mathrm{int}}$.

An orientable object is characterized by $10$ quantum numbers. $8$ of them
have a standard interpretation (these are the $4$-momentum $p^{\mu }$, spin
$s$, helicity, and representation $(j_{1},j_{2})$ of the Lorentz group).

Two additional quantum numbers $S_{3}^{R}$ and $B_{3}^{R}$ that correspond
to the generators $\hat{S}_{3}^{R}$ and $\hat{B}_{3}^{R}$
of the group $SL(2,C)_{\mathrm{int}}\subset M(3,1)_{\mathrm{int}}$ can be
interpreted as some charges. Indeed, the charges are additive quantum
numbers, being independent of a choice of the laboratory reference frame.
Generators of $SL(2,C)_{\mathrm{int}}$ commute with the generators
$M(3,1)_{\mathrm{ext}}$ (i.e., with the generators of the Lorentz transformations),
and therefore \textquotedblleft right\textquotedblright\ quantum numbers
$S_{3}^{R}$ and $B_{3}^{R}$ do not change under a change
of the laboratory reference frame.

The two additional \textquotedblleft right\textquotedblright\ quantum
numbers\ $S_{3}^{R}$ and $B_{3}^{R}$ characterizing
orientable objects possess some properties with respect to discrete
transformations, and their possible values are related to the spin value. It
was noted that \textquotedblleft right\textquotedblright\ quantum numbers
can be (in fact, uniquely) ascribed to all known elementary particles. Thus,
we believe that the complete and, therefore, more adequate description of
elementary (spinning) particles is achieved if one considers them as
orientable objects and use the corresponding relativistic classification
theory developed in this work.

In spite of the fact that left and right transformations commute, the
spectra of left and right generators $\hat{S}_{3}^{R}$ and $\hat{B}_{3}^{R}$
are not independent. In particular, the \textquotedblleft
right\textquotedblright\ charges $S_{3}^{R}$ and $B_{3}^{R}$ must be integer
for particles with integer spin and half-integer for particles with
half-integer spin. Note that, if $S_{3}^{R}$ supposed to be a conserved
quantum number, then $B_{3}^{R}$ has a definite value only for states with a
definite chirality (but not parity, since $\hat{B}_{3}^{R}$ does not commute
with the parity operator), i.e., $B_{3}^{R}$ can be conserved only for
massless particles.

The classification of orientable objects yields the following properties in
the one-particle sector.
For fermions of spin $1/2$, there are four states (quadruplet, realized by
the up/down components of a weak doublet and by their antiparticles),
distinguished by right generators of the Poincar\'{e} group: the sign of
$B_{3}^{R}$-charge (instead of which one can choose the sign of chirality or
internal parity) and by the sign of $S_{3}^{R}$-charge.
Particles of spin $1$ also form a quadruplet, whose quantum numbers coincide
with those of $W^{+},W^{+},W^{0},A^{0}$. Besides, potentially there are
another $6$ states corresponding to the representations
$T_{[10]}\oplus T_{[01]}$ of the group $SL(2,C)_{\mathrm{int}}$.

Note, once again, that in contrast with the left (external) symmetries, the
right (internal) symmetries can be generally broken, and, respectively,
states related by these symmetries can have different characteristics
(including mass).

We have considered a description on a basis of finite-dimensional
representations of the Lorentz group (and the related finite-dimensional
RWE). In our following work, we hope to consider unitary
infinite-dimensional representations of the Lorentz group and the related
equations of the Majorana type, as well as the case of interaction.

\bigskip \textbf{Acknowledgements}\quad
D.M.G. acknowledges the permanent support of FAPESP and CNPq.

%%%%%%%%%%%%%%%%%%%%%%%%%%%%%%%%%%%%%%%%%%%%%%%%%%%%%%%%%%%%%

\section*{Appendix. Generators and weight diagrams of the Lorentz group}

Besides the four-dimensional vector notation for spin operators (see (\ref{gen.SL}),(\ref{gen.SR})),
it is also convenient to use a three-dimensional notation:
$\hat{S}_{k}=\frac{1}{2}\epsilon _{ijk}\hat{S}^{ij}$, $\hat{B}_{k}=\hat{S}_{0k}$.
In the space of functions on the group $f(z^{\alpha },\,\uz^{\alpha })$
(functions of the $4$ elements of a matrix $SL(2,C)$ (\ref{par.0c}))
a direct calculation yields for left and right generators{%
\footnote{%
For the sake of brevity, we have used the notation that we applied in \cite{GitSh09,GitSh01},
$z^{\alpha }=z_{\;\;\underline{1}}^{\alpha }$, $\cc z_{\dot{\alpha}}=\cc z_{\dot{\alpha}}^{\;\;\dot{\underline{2}}}$,
${\underline{z}}^{\alpha }=z_{\;\;\underline{2}}^{\alpha }$,
$\cc {\underline z}_{\dot{\alpha}}=\cc z_{\dot{\alpha}}^{\;\;\dot{\underline{1}}}$.}}
\cite{GitSh01}
\begin{eqnarray}
\label{SL} &&\hat S_k=\frac 12 (z\sigma_k\partial _z -\cc
z\cc\sigma_k\partial _{\ccc z}\,)+... \; ,
\nonumber \\
&&\hat B_k=
\frac i2 (z\sigma_k\partial _z + \cc z\cc\sigma_k\partial _{\ccc z}\,)+... \; ,
\quad z=(z^1\; z^2), \quad
\partial_z=(\partial/\partial{z^1}\; \partial/\partial{z^2})^T ;
\\   \label{SR}
&&\hat S_k^R=-\frac 12 (\chi\cc\sigma_k\partial_\chi
-\cc\chi\sigma_k\partial _{\ccc \chi}\,)+... \; ,
\nonumber \\
&&\hat B_k^R=
-\frac i2 (\chi\cc\sigma_k\partial_\chi + \cc\chi\sigma_k\partial _{\ccc \chi}\,)+... \; ,
\quad \chi=(z^1\;\uz^1), \quad
\partial_\chi=(\partial/\partial{z^1}\; \partial/\partial{\uz^1})^T ;
\end{eqnarray}
The terms $...$ stand for analogous expressions obtained by the change
$z\to z'=(\uz^1\;\uz^2)$, $\chi\to \chi'=(z^2\; \uz^2)$. In particular,
\begin{equation}
\hat S_3^R= \frac 12 (-z\partial _z + \uz\partial _\uz + \cc z\partial _{\ccc z} - \cc \uz\partial _{\ccc \uz}\,) \; ,\qquad
\hat B_3^R= \frac i2 (-z\partial _z + \uz\partial _\uz - \cc z\partial _{\ccc z} + \cc \uz\partial _{\ccc \uz}\,) \; .
\label{S3B3}
\end{equation}

It is known that from $\hat S_k$ and $\hat B_k$ one can construct
linear combinations $\hat M_k$ and $\barM_k$,
\begin{eqnarray}
&&\hat M_k=\frac 12(\hat S_k-i\hat B_k)=z\sigma_k\partial _z + \uz\sigma_k\partial _\uz, \quad
\hat M_+=z^1\partial /\partial{z^2}, \quad
\hat M_-=z^2\partial /\partial{z^1},
\nonumber \\
&&\barM_k=-\frac 12(\hat S_k+i\hat B_k)=\cc z\cc\sigma_k\partial _{\ccc z} + \cc\uz\cc\sigma_k\partial _{\ccc\uz}\,, \quad
\barM_+=\cc\uz^{\dot 1}\partial /\partial{\cc\uz^{\dot 2}}, \quad
\barM_-=\cc\uz^{\dot 2}\partial /\partial{\cc\uz^{\dot 1}},
\label{MNgen}
\end{eqnarray}
such that $[\hat M_i,\barM_k]=0$; in addition, for unitary
representations of the Lorentz group, as it follows from the condition
$\hat S_k^\dagger =\hat S_k$, $\hat B_k^\dagger =\hat B_k$, the relation
$\hat M_k^\dagger =\barM_k$ must be fulfilled (for finite-dimensional non-unitary
representations $\hat S_k^\dagger =\hat S_k$, $\hat B_k^\dagger =-\hat B_k$ and
$\hat M_k^\dagger =-\barM_k$).

Taking into account the fact that the operators $\hat M_k$ and $\barM_k$
satisfy commutation relations of the algebra $su(2)$,
we find the following relations for the spectra of the Casimir operators of
the Lorentz subgroup:
\begin{eqnarray}
&&\hat {\mathbf S}^2 -\hat {\mathbf B}^2 = 2(\hat {\mathbf M}^2+\barMM)
  =2j_1(j_1+1)+2j_2(j_2+1)=-\frac 12 (k^2-\rho ^2 -4), \quad
\nonumber \\
&&\hat {\mathbf S}\hat {\mathbf B} = -i(\hat {\mathbf M}^2-\barMM)
  =-i\left( j_1(j_1+1)-j_2(j_2+1)\right) =k\rho,\qquad
\nonumber \\
&&\hbox{where } \quad \rho=-i(j_1+j_2+1), \quad  k=j_1-j_2.
\label{Lcas1}
\end{eqnarray}
That is, the irreps of the Lorentz group $SL(2,C)$ are labeled by a pair of
numbers $[j_1,j_2]$. It is convenient to label unitary infinite-dimensional
irreps by pairs of numbers $(k,\rho)$; in addition, the irreps $(k,\rho)$
and $(-k,-\rho)$ are equivalent \cite{BarRa77,GelGrV66}.

For finite-dimensional and unitary infinite-dimensional irreps of the group
$SL(2,C)$, the formulae of reduction to the compact $SU(2)$-subgroup have the
respective form
\begin{equation}  \label{reduc1}
T_{[j_1,j_2]}=\sum_{j=|j_1-j_2|}^{j_1+j_2}T_j, \qquad
T_{(k,\rho)}=\sum_{j=k}^{\infty}T_j,
\end{equation}
see \cite{GelGrV66}.

The difference $j_1-j_2$ (the difference between the number of dotted and
undotted indices) can also be obtained as an eigenvalue of the chirality
operator $\hat \Gamma^5$ (\ref{chir}).

Representations of low dimensions have a simple realization. The
two-dimensional irreps $T_{[1/2\; 0]}$ and $T_{[0\; 1/2]}$, which induce the
transformations of spinors -- these are complex-conjugate matrices from
SL(2,C), three-dimensional irreps $T_{[1\; 0]}$ and $T_{[0\; 1]}$ --
complex-conjugate matrices from $SO(3,C)$, and four-dimensional matrices $%
T_{[1/2\; 1/2]}$, which induce the transformations of 4-vectors -- this is a
representation by real-valued matrices from $SO(3,1)$.

The weight diagrams of the representations $T_{[1/2\; 0]}\oplus T_{[0\;
1/2]} $ and $T_{[1\; 0]}\oplus T_{[0\; 1]}$ are given by the figure. In
addition, the axes $S_3$ and $-iB_3$ on which we indicate the eigenvalues of
the corresponding operators are rotated by the angle of $45^\circ$ with
respect to the axes $m_1$ and $m_2$, on which we indicate eigenvalues of the
operators $\hat M_3$ and $\barM_k$.

\bigskip
\unitlength=1.1pt
\linethickness{0.4pt}

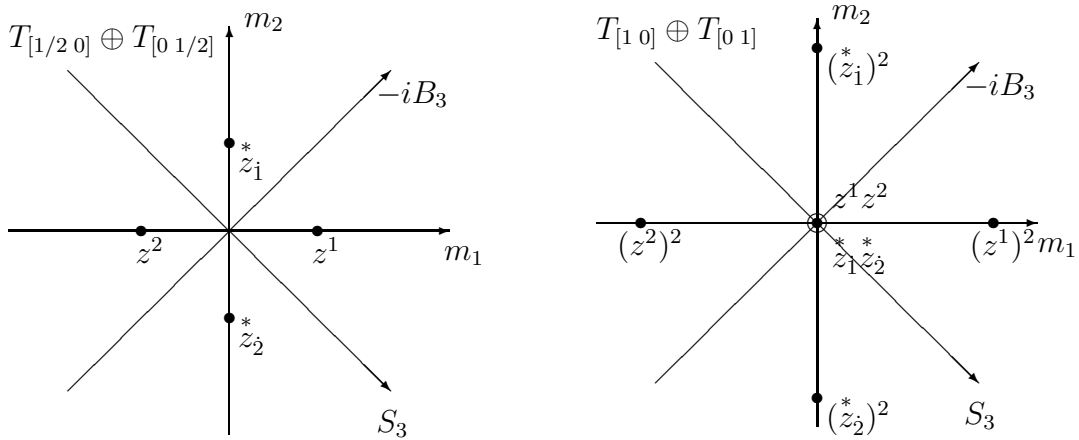
\newsavebox{\spinone}
\savebox{\spinone}(160,150)[lb]
{
\put(0,70){\vector (1,0){150}}
\put(75,0){\vector (0,1){140}}
\put(20,125){\vector (1,-1){110}}
\put(20,15){\vector (1,1){110}}
\put(0,133){$T_{[1\; 0]}\oplus T_{[0\; 1]}$}
\put(125,1){$S_3$}
\put(125,114){$-iB_3$}
\put(75,10){\circle*{4}}
\put(75,130){\circle*{4}}
\put(78,0){$(\cc z_{\dot 2})^2$}
\put(78,120){$(\cc z_{\dot 1})^2$}
\put(15,70){\circle*{4}}
\put(7,59){$(z^2)^2$}
\put(127,59){$(z^1)^2$}
\put(135,70){\circle*{4}}
\put(75,70){\circle*{4}}
\put(75,70){\circle{6}}
\put(80,75){$z^1z^2$}
\put(80,55){$\cc z_{\dot 1}\cc z_{\dot 2}$}
\put(150,60){$m_1$}
\put(80,140){$m_2$}
}
\begin{picture}(360,150)
\put(0,70){\vector (1,0){150}}
\put(75,0){\vector (0,1){140}}
\put(20,125){\vector (1,-1){110}}
\put(20,15){\vector (1,1){110}}
\put(0,133){$T_{[1/2\; 0]}\oplus T_{[0\; 1/2]}$}
\put(125,1){$S_3$}
\put(125,114){$-iB_3$}
\put(75,40){\circle*{4}}
\put(78,30){$\cc z_{\dot 2}$}
\put(75,100){\circle*{4}}
\put(78,90){$\cc z_{\dot 1}$}
\put(45,70){\circle*{4}}
\put(43,59){$z^2$}
\put(105,70){\circle*{4}}
\put(103,59){$z^1$}
\put(148,60){$m_1$}
\put(80,140){$m_2$}

\put(200,0){\usebox{\spinone}}

\end{picture}

%\markright{References}
%\bibliography{abrv,book,group,releq,quant,book_rus,probab,path,gauge}

\begin{thebibliography}{99}
\bibitem{GitSh09}
D.M. Gitman and A.L. Shelepin.
\newblock Fields on the {P}oincar\'e group and quantum description of
  orientable objects.
\newblock {\em Eur. Phys. J. C}, 61(1):111--139, 2009.
\newblock arXiv:hep-th/0901.2537.

\bibitem{Wig39}
E.P. Wigner.
\newblock On unitary representations of the inhomogeneous {L}orentz group.
\newblock {\em Ann. Math.}, 40(1):149--204, 1939.

\bibitem{ColMa67}
S.~Coleman and J.~Mandula.
\newblock All possible symmetries of the {$S$}-matrix.
\newblock {\em Phys. Rev.}, 159(5):1251--1256, 1967.

\bibitem{BarRa77}
A.O. Barut and R.~Raczka.
\newblock {\em Theory of Group Representations and Applications}.
\newblock PWN, Warszawa, 1977.

\bibitem{Wigne59}
E.P. Wigner.
\newblock {\em Group Theory and its Application to the Quantum Mechanics of
  Atomic Spectra}.
\newblock Academic Press, New York, 1959.

\bibitem{LanLi3}
L.D. Landau and E.M. Lifschitz.
\newblock {\em Quantum Mechanics}, volume 3 of Course of Theoretical Physics.
\newblock Pergamon, Oxford, 1977.

\bibitem{BieLo81}
L.S. Biedenharn and J.D. Louck.
\newblock {\em Angular Momentum in Quantum Physics}.
\newblock Addison-Wesley, Reading, Massachusetts, 1981.

\bibitem{Zare88}
R.N. Zare.
\newblock {\em Angular Momentum. Understanding Spatial Aspects in Chemistry and
  Physics}.
\newblock Wiley, New York, 1988.

\bibitem{LurMi61}
F.~Lur{\c c}at and L.~Michel.
\newblock Sur les relations entre charges et spin.
\newblock {\em Nuovo Cimento}, 21(3):574--576, 1961.

\bibitem{KayGo83}
B.~Kayser and A.S. Goldhaber.
\newblock {CPT} and {CP} properties of {M}ajorana particles, and the
  consequences.
\newblock {\em Phys. Rev. D}, 28(9):2341--2344, 1983.

\bibitem{KayGiP89}
B.~Kayser, F.~Gibrat-Debu, and F.~Perrier.
\newblock {\em The Physics of Massive Neutrinos}.
\newblock World Scientific, Singapore, 1989.

\bibitem{GitSh01}
D.M. Gitman and A.L. Shelepin.
\newblock Fields on the {P}oincar\'e group: Arbitrary spin description and
  relativistic wave equations.
\newblock {\em Int. J. Theor. Phys.}, 40:603--684, 2001.
\newblock arXiv:hep-th/0003146.

\bibitem{BucGiS02}
I.L. Buchbinder, D.M. Gitman, and A.L. Shelepin.
\newblock Discrete symmetries as automorphisms of the proper {P}oincar\'e
  group.
\newblock {\em Int. J. Theor. Phys.}, 41(4):753--790, 2002.
\newblock arXiv:hep-th/0010035.

\bibitem{BerLiP}
V.B. Berestetskii, E.M. Lifshitz, and L.P. Pitaevskii.
\newblock {\em Relativistic Quantum Theory}.
\newblock Pergamon, New York, 1971.

\bibitem{Wig63}
E.P. Wigner.
\newblock In A.~Salam, editor, {\em Theoretical Physics}, page~59, Trieste,
  1963. IAEA.

\bibitem{KimWi87}
Y.S. Kim and E.P. Wigner.
\newblock Cylindrical group and massless particles.
\newblock {\em J. Math. Phys.}, 28(5):1175--1179, 1987.

\bibitem{BieBrT88}
L.C. Biedenharn, H.W. Braden, P.~Truini, and H~{Van Dam}.
\newblock Relativistic wavefunctions on spinor spaces.
\newblock {\em J. Phys. A}, 21:3593--3610, 1988.

\bibitem{HasSi92}
Z.~Hasiewicz and P.~Siemion.
\newblock A bosonic model for particles with arbitrary spin.
\newblock {\em Int. J. Mod. Phys. A}, 7(17):3979--3996, 1992.

\bibitem{KuzLy95}
S.M. Kuzenko, S.L. Lyakhovich, and A.Yu. Segal.
\newblock A geometric model of the arbitrary spin massive particle.
\newblock {\em Int. J. Mod. Phys. A}, 10(10):1529--1552, 1995.

\bibitem{LyaSeS96}
S.L. Lyakhovich, A.Yu. Segal, and A.A. Sharapov.
\newblock Universal model of a {$D=4$} spining particle.
\newblock {\em Phys. Rev. D}, 54(8):5223--5238, 1996.

\bibitem{BacKi69}
H.~Bacry and A.~Kihlberg.
\newblock Wavefunctions on homogeneous spaces.
\newblock {\em J. Math. Phys.}, 10(12):2132--2141, 1969.

\bibitem{Kihlb70}
A.~Kihlberg.
\newblock Fields on a homogeneous space of the {P}oincare group.
\newblock {\em Ann. Inst. Henri Poincar{\'e}}, 13(1):57--76, 1970.

\bibitem{BoyFl74}
C.P. Boyer and G.N. Fleming.
\newblock Quantum field theory on a seven-dimensional homogeneous space of the
  {P}oincar\'e group.
\newblock {\em J. Math. Phys.}, 15(7):1007--1024, 1974.

\bibitem{DerGi99}
A.A. Deriglazov and D.M. Gitman.
\newblock Classical description of spinning degrees of freedom of relativistic
  particles by means of commuting spinors.
\newblock {\em Mod. Phys. Lett. A}, 14:709--720, 1999.

\bibitem{Drech97}
W.~Drechsler.
\newblock Geometro-stohastically quantized fields with internal spin variables.
\newblock {\em J. Math. Phys.}, 38(11):5531--5558, 1997.

\bibitem{Varla05}
V.V. Varlamov.
\newblock {M}axwell field on the {P}oincare group.
\newblock {\em Int. J. Mod. Phys. A}, 20:4095--4112, 2005.
\newblock arXiv:math-ph/0310051.

\bibitem{Finke55}
D.~Finkelstein.
\newblock Internal structure of spinning particles.
\newblock {\em Phys. Rev.}, 100(3):924--931, 1955.

\bibitem{GelGrV66}
I.M. Gel'fand, M.I. Graev, and N.Ya. Vilenkin.
\newblock {\em Generalized Functions}, volume~5.
\newblock Academic Press, New York, 1966.

\end{thebibliography}
%\bibliographystyle{unsrt}

\end{document}